\documentclass[prd,twocolumn,twoside,showpacs,superscriptaddress]{revtex4}
\usepackage{graphics,amssymb,amsmath,epsf,graphicx}
\epsfclipon

\usepackage{xspace}

\begin{document}

\title{An Infrared Safe perturbative approach to Yang-Mills correlators}

\author{Matthieu Tissier}
\affiliation{Laboratoire de Physique Th\'eorique de la Mati\`ere Condens\'ee,
Universit\'e Pierre et Marie Curie, 4 Place Jussieu 75252 Paris CEDEX 05, France}
\author{Nicol\'as Wschebor} 
\affiliation{Instituto de F\'{\i}sica, Facultad de Ingenier\'{\i}a, Universidad de la Rep\'ublica, 
J.H.y Reissig 565, 11000 Montevideo, Uruguay}

\begin{abstract}
We investigate the 2-point correlation functions of Yang-Mills theory
in the Landau gauge by means of a massive extension of the
Faddeev-Popov action. This model is based on some phenomenological
arguments and constraints on the ultraviolet behavior of the
theory. We show that the running coupling constant remains finite at
all energy scales (no Landau pole) for $d>2$ and argue that the relevant
parameter of perturbation theory is significantly smaller than 1 at
all energies. Perturbative results at low orders are therefore
expected to be satisfactory and we indeed find a very good agreement
between 1-loop correlation functions and the lattice simulations, in 3
and 4 dimensions. Dimension 2 is shown to play the role of an upper
critical dimension, which explains why the lattice predictions are
qualitatively different from those in higher dimensions.
\end{abstract}
\date{\today}
\maketitle

\section{Introduction}
There is up to day no fully satisfying covariant gauge fixing of
non-abelian gauge theories. The common Faddeev-Popov (FP) procedure is
known to be invalid because of the Gribov ambiguity \cite{Gribov77}:
the gauge constraint (such as $\partial_\mu A_\mu^a=0$ in the Landau
gauge) have many solutions (Gribov copies) equivalent up to a gauge
transformation. Taking into account this ambiguity would give
non-perturbative contributions (typically of the form
$\exp(-\text{cst.}/g^2)$) that vanish at all orders of the
perturbation theory. Consequently, the FP procedure is actually
justified when studying high energy phenomena and the perturbative
predictions are in excellent agreement with the experiments. On the
other range of the spectrum, when considering infrared (IR)
properties, the perturbative analysis extracted from the FP procedure
is inconsistent because the coupling becomes large. The coupling is
found to diverge at a finite energy scale (known as a Landau pole) but
of course this prediction is out of the range of validity of
perturbation theory. There are two possible explanations of the
failure of the perturbative predictions of the FP procedure. Either
the coupling indeed reaches large values (but this seems to be at odds
with the lattice simulation results) or some nonperturbative effects,
such as the Gribov ambiguity, invalidates the FP procedure itself.

In fact, we can avoid fixing the gauge by simulating gauge invariant
quantities on the lattice and most of the IR studies are done in this
way. However it would be convenient to have some analytic (or
semi-analytic) predictions in that regime, which seem to require to
fix the gauge (see however \cite{Morris99,Morris00,Morris05}), and
therefore to take into account the Gribov ambiguity.

Several analytic methods have been considered in order to access the
IR properties. The most developed is based on the Schwinger-Dyson
(SD)
equations \cite{vonSmekal97,Alkofer00,Zwanziger01,Fischer03,Bloch03,Aguilar04,Boucaud06,Aguilar07,Aguilar08,Boucaud08,Fischer08,RodriguezQuintero10},
which consist in an infinite set of coupled equations for the vertex
functions. In order to make predictions, it is necessary to truncate
in some way this infinite set. Different schemes have been
proposed but most of them consider the equations for the 2-point
functions with an ansatz for the 3 and 4-point functions. (Most of the
analyses have been done in the Landau gauge and we restrict the
discussion to this gauge in the rest of the paper.) This leads to some
predictions on the behavior of the ghost and gluon propagators and two
types of solutions have been found. In the so-called scaling
solution \cite{vonSmekal97,Alkofer00,Zwanziger01,Fischer03,Bloch03,Fischer08},
both propagators have a power-law behavior in the IR, the ghost
propagator is more singular than the bare one and the gluon propagator
approaches zero in that limit. The exponents governing these two
power-laws are not independent. In the so called decoupling solution
(or massive solution)
 \cite{Bloch03,Aguilar04,Boucaud06,Aguilar07,Aguilar08,Boucaud08,RodriguezQuintero10},
the gluon propagator goes to a constant in the IR and the ghost
propagator is as singular as in the bare theory.

These correlation functions were also studied in the framework of the
Non-perturbative Renormalization Group (NPRG)
equation \cite{Wetterich92,Berges00}. Again, one has to make some
truncations in an infinite set of coupled
equations \cite{Pawlowski03,Fischer04,Fischer08}. The results again
depend on the approximation scheme, but seem to be consistent with the
scaling solution.

A third approach, known as the Gribov-Zwanziger (GZ)
model \cite{Gribov77,Zwanziger89,Zwanziger92}, relies more specifically
on the influence of the Gribov copies. By introducing several
auxiliary fields, it is possible to restrict effectively the
functional integration on the gluon field to the first Gribov region
where the FP operator is positive definite. It was originally expected
that this restriction would give an unambiguous gauge fixing but it
was later realized that many copies of some field configuration were
present within the first Gribov region \cite{Semenov86}. This procedure significantly
reduces the number of Gribov copies but unfortunately not to a single
one. The Gribov ambiguity is therefore still present in this model. In
the first implementations of the GZ model, the predictions for the
propagators were consistent with the scaling solution. More recently
a refined version of this model was introduced \cite{Dudal08}, that
takes into account the appearance of some condensates. The results are
then consistent with the massive solution of SD equations.

The results obtained in the SD approach triggered a large activity
from the lattice community. Since the predictions of the SD approaches
are mostly for gauge dependent quantities, it was necessary to
implement a gauge fixing procedure in the lattice, at odds with usual
simulations. One of the merits of the precursor work of Alkofer and
von~Smekal \cite{vonSmekal97,Alkofer00} was to discard very singular
solutions found previously for the gluon propagator and this was
confirmed by the lattice simulations. However, a broad consensus on
the details of the simulation results in $d=3$ and 4 was achieved only
recently, mainly through large lattice simulations
\cite{Cucchieri_08b,Cucchieri_08,Cucchieri09,Bogolubsky09,Dudal10}. It
is now well established that in these dimensions, the simulations
support the massive solution. The $d=2$ case is different in this
respect~\cite{Cucchieri_08,Maas_07} and is well described by the
scaling solution and the explanation of this difference is still
unclear (see however \cite{Dudal08b}).

Recently \cite{Tissier10}, we proposed an alternative to the
analytical methods described above. Its motivation relies on the fact
that we do not have so far a fully justified covariant gauge fixing
procedure that can be handled in analytical calculations. Ideally, we
should try to construct such a unambiguous gauge-fixing procedure, but
this is an extremely hard task. Alternatively, we propose to construct
a model that uses as guiding principle the behavior of correlation
functions observed on the lattice and that respects as many important
known properties of correlation functions as possible. More precisely,
since the gluon propagator is observed to be massive in the lattice
simulations (for $d>2$), we propose to include this mass term directly
at the bare level. (This corresponds to a particular case of the
Curci-Ferrari model \cite{Curci76}.) This modifies the theory in the
IR but preserves the standard FP predictions for momenta $p\gg m$ at
all orders of perturbation theory. In particular, the mass term does not spoil the renormalizability.
This model reproduces at one loop
order with excellent precision the lattice predictions. We also showed
that the spectral function of the gluons is not positive-definite, in
agreement with other studies \cite{Bowman07,Cucchieri_04,Alkofer00}.

Our aims in the present article are the following. First, we give a
detailed version of the calculations presented in \cite{Tissier10},
including some other one loop calculations that were not presented in
the previous work for lack of space. Second, we show that an
appropriate renormalization scheme can be chosen that does not present
a Landau pole in the IR. Third, we discuss in detail the size of
higher loop corrections showing that, in fact, they seem to be rather
small. This opens the door to perturbative calculations in QCD in the
IR regime, with a huge number of possible applications. Fourth, we
analyze the $d=2$ case and explain why in the present model it is very
different from the $d>2$ case.

The article is organized as follows. In Sect. II, we describe in
detail the model and its general properties. In Sect. III, we present
the 1-loop perturbative calculation of the 2-point functions and
compare our results with the lattice data in $d=4$ and $d=3$. In
Sect. IV, we perform a renormalization group (RG) analysis of these
propagators. We propose two different renormalization schemes, one of
which is shown to be IR safe in the sense that it leads to no Landau
pole in the IR. We compare the RG results with the lattice data in
$d=4$ and $d=3$. In Sect. V, we discuss the $d=2$ case. Some technical
details are presented in two appendices.

\section{The model}
\label{themodel}

As said in the Introduction, the FP action is not justified at a
non-perturbative level. In this section, we present a modification of
the FP action in the Landau gauge, based on phenomenological
considerations.  Our main guide is the observation that the gluon
propagator tends to a finite positive value in the IR for $d>2$. We
propose to impose this property at the tree level by adding a mass
term for the gluon in the FP action.  We do not change the ghost
sector since the ghost propagator is found to be IR divergent in the
simulations. This may also be motivated by assuming that the shift
symmetry $\bar c\to \bar c + cst.$ is
preserved beyond perturbation theory.
Moreover, we do not want to modify interactions in the action so as to
preserve the ultraviolet (UV) behavior of the theory and maintain the
predictions of perturbative QCD (or gluodynamics) for momenta much
larger than $\Lambda_{QCD}$.  From this analysis it is clear that if
we choose not to modify the field content of the theory, mass terms
are the only local and renormalizable modifications of the FP action
that do not affect the UV behavior of the model. If we restrict to
local terms, the only other possible way to modify the action is to
introduce new fields as done in the GZ
model~\cite{Gribov77,Zwanziger89,Zwanziger92}. The origin of the
appearance of the effective mass term in the IR is a complex problem
and could be related with the Gribov ambiguity, with some kind of
condensates (see for example \cite{Verschelde01,Browne03,Dudal08}), or with other
nonperturbative effects, but our phenomenological approach does not
rely on which of these scenarios is valid.

This analysis leads us to consider the
Landau-gauge FP euclidean Lagrangian for pure gluodynamics, supplemented with a gluon mass
term:
\begin{equation}
  \label{eq_lagrang}
  \mathcal L=\frac 14 (F_{\mu\nu}^a)^2+\partial _\mu\overline c^a(D_\mu
  c)^a+ih^a\partial_\mu A_\mu^a+\frac {m^2}2 (A_\mu^a)^2 ,
\end{equation}
where $(D_\mu c)^a=\partial_\mu c^a+g f^{abc}A_\mu^b c^c$, the
field strength $F_{\mu\nu}^a=\partial_\mu A_\nu^a-\partial_\nu
A_\mu^a+gf^{abc}A_\mu^bA_\nu^c$ are expressed in terms of the coupling
constant $g$ and the latin indices correspond to the $SU(N)$ gauge group. The Lagrangian~(\ref{eq_lagrang}) corresponds to a
particular case of the Curci-Ferrari model~\cite{Curci76}. At the tree
level, the gluon propagator is massive and transverse in momentum
space:
\begin{equation}
\label{eq_propag_bare}
  G_{\mu\nu}^{ab}(p)=\delta^{ab} P^\perp_{\mu\nu}(p)\frac 1{p^2+m^2},
\end{equation}
with $P^\perp_{\mu\nu}(p)=\delta_{\mu\nu}-p_\mu p_\nu/p^2$.  It is
interesting to note that the spectral density associated with the
propagator~(\ref{eq_propag_bare}) is positive and therefore there is
no violation of positivity at the tree level. As discussed in
\cite{Tissier10}, violations of positivity are present in the model but
they are caused by fluctuations.

The gluon propagator observed in the lattice is not
compatible at a quantitative level with the bare propagator~(\ref{eq_propag_bare}) and we will
show below that, by including the one-loop corrections, one obtains
propagators for gluons and ghosts that are in impressive agreement
with those obtained in the lattice in $d=4$ and $d=3$.

Let us mention that a mass term has been used to improve perturbative
QCD results in order to reproduce the phenomenology of Strong
Interactions~\cite{Parisi80,Cornwall82,Natale09}. Moreover, there are successful
confinement models~\cite{Cornwall79} that use actions including a
gluon mass term. The difference with the model used in those works is
that the Curci-Ferrari model is renormalizable, allowing to perform
perturbative calculations at any order. This also implies that the UV
beta function identifies with the standard results of the FP
procedure.

When analyzing the model described above, we must face the problem
that the mass term breaks the BRST symmetry~\cite{BRS,T} which is very
important in the perturbative analysis.  This symmetry has the form:
\begin{equation}
\label{BRST}
\begin{array}{ll}
  \delta A_{\mu}^a= \eta\, (D_\mu c)^a, &\delta c^a= - \eta\,\frac g 2 f^{abc} c^b c^c, \\
  \delta \bar c^a= \eta\, i h^a, &\delta i h^a = 0,
\end{array}
\end{equation}
where $\eta$ is a global grassmanian parameter.  The BRST symmetry is
in general used to prove the renormalizability of the theory. However,
the breaking of the BRST symmetry by the mass term is soft and
therefore does not spoil renormalizability~\cite{Curci76,deBoer95}.

The BRST symmetry is also used to reduce the state space to the
physical space, in which the theory is unitary (at least at the
perturbative level) and breaking this symmetry spoils the standard
proof of unitarity. This problem is actually common to essentially all
methods that try to go beyond the standard perturbation theory (as the
GZ model) because they all break the standard BRST symmetry. In this
respect, the model considered here is not in a worse position than
other approaches considered in the field. We must insist that this
model is equivalent to the standard FP model in the UV limit $p \gg
\Lambda_{QCD}$ if $m \sim \Lambda_{QCD}$. This means that in the
domain of validity of standard perturbation theory, the model is as
unitary as QCD. The unitarity of the model in other momentum regimes
is of course an important open problem, as it is in all gauge fixings
in which standard BRST symmetry is broken.

The model with Lagrangian~(\ref{eq_lagrang}), as a particular case of
the Curci-Ferrari model, has a pseudo-BRST symmetry (not nilpotent)
that has the same form as the standard BRST~(\ref{BRST}) except for
the $h$ variation which reads $\delta i h^a = \eta\, m^2 c^a$. On top
of this symmetry, the Lagrangian has all the standard symmetries of
the FP action for the Landau gauge. This includes the shift in
antighost $\bar c\to \bar c +cst.$, a symplectic
group~\cite{Delduc89}, and four gauged supersymmetries \footnote{In
  the Ref. \cite{Tissier08} a different but equivalent form of the
  action has been employed that makes manifest the symmetry $c
  \leftrightarrow \bar c$.} ~\cite{Tissier08} .  As a consequence, the
mass~\cite{Dudal02} and coupling constant~\cite{Taylor71}
renormalization factors are fixed in terms of gluon and ghost field
renormalizations. More precisely, by using the standard definition of
renormalization factors (here the subscript 'B' denote bare
quantities corresponding to their respective renormalized quantities
without subscript):
\begin{align}
 A_B^{a\,\mu}= \sqrt{Z_A} A^{a\,\mu},&\hspace{.5cm} 
 c_B^{a}= \sqrt{Z_c} c^{a},\hspace{.5cm}
 \bar c_B^{a}= \sqrt{Z_c} \bar c^{a},\hspace{.5cm} \nonumber\\
g_B&= Z_g g \hspace{.5cm} m_B^2= Z_{m^2} m^2
\end{align}
one can prove that the divergent part of the renormalization factors
(or, similarly, the renormalization factors themselves in a
$\overline{MS}$ scheme) verify the two non-renormalization theorems
\cite{Taylor71,Dudal02}:
\begin{align}
\label{no-renorm1}
Z_g \sqrt{Z_A} Z_c &=1,\\
\label{no-renorm2}Z_{m^2} Z_A Z_c &=1.
\end{align}
These non-renormalization theorems are particular cases of those found
in the Curci-Ferrari model for any gauge parameter $\xi$, as proven
recently \cite{Wschebor07,Tissier08}.  The finite parts of
renormalized vertices also verify non-renormalization theorems, as
will be discussed below, but their explicit form depend on the
considered renormalization scheme.

\section{Perturbative calculation of 2-point correlation functions}
\label{strictperturb}
\begin{figure}[tbp]
\includegraphics[width=2.5cm]{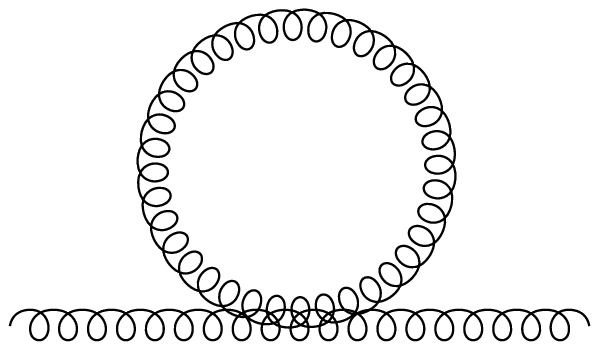}
\includegraphics[width=2.5cm]{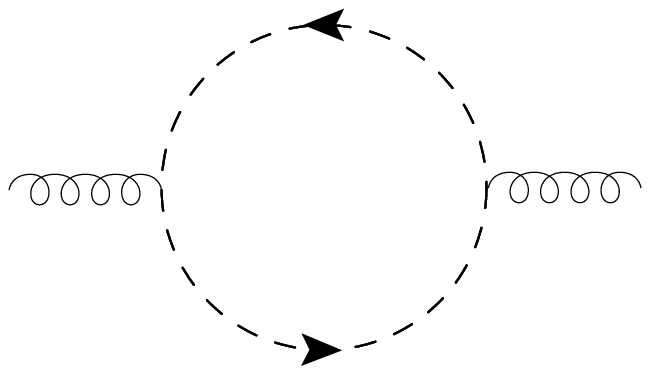}
\includegraphics[width=2.5cm]{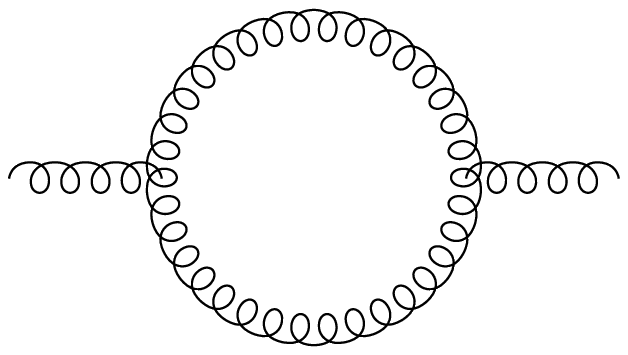}

\vspace{.8cm}
\includegraphics[width=2.5cm,angle=0]{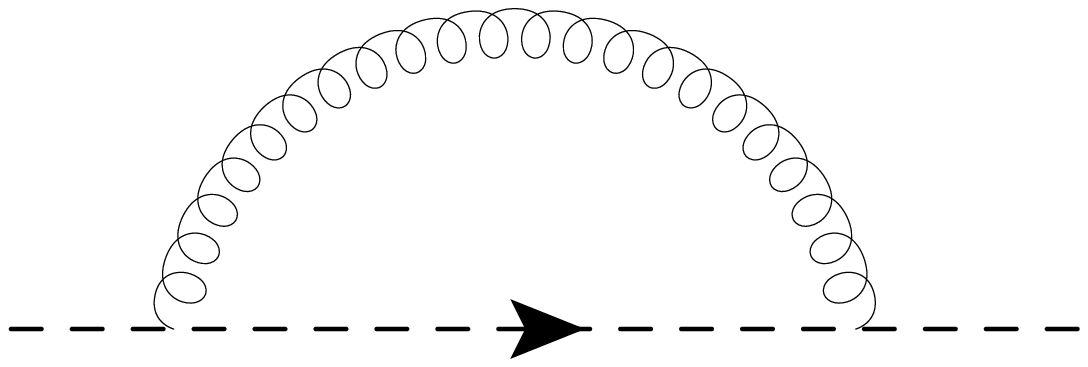}
\caption{\label{diagrams}First line: three diagrams contributing to the gluon self-energy. Second line: diagram contributing
to the ghost self-energy}
\end{figure}

We present in this section a strict 1-loop calculation (that is,
without taking into account RG effects) of gluon and ghost propagators
in arbitrary dimension.  These correlators require the calculation of
four Feynman diagrams as shown in Fig.~\ref{diagrams}.  
The
corresponding results in $d=4$ and $d=3$ are then compared with
lattice results, showing a very good agreement (for momenta not much
larger than $m$).  Some of these results were presented in
\cite{Tissier10} but for completeness we review all of them here.

Observing that the gluon propagator is transverse in Landau gauge, it
is convenient to parametrize the gluon $G_{\mu\nu}^{ab}(p)$ and the
ghost $G^{ab}(p)$ propagators in the form:
\begin{equation}
G^{ab}(p)=\delta_{ab}F(p)/p^2, \hspace{.2cm}
G_{\mu\nu}^{ab}(p)=P^\perp_{\mu\nu}(p) \delta_{ab} G(p).
\end{equation}
The $F(p)$ is known as the ghost dressing function and the scalar
function $G(p)$ will be referred to as the gluon propagator below. We also define the 2-point vertex functions
\begin{align}
\Gamma_A^{(2)}(p)&=G^{-1}(p), \nonumber\\
\Gamma^{(2)}_{\bar c c}(p)&=p^2 F^{-1}(p).
\end{align}
We choose the following renormalization conditions for 2-point functions:
\begin{align}
\label{rencond}
&\Gamma_A^{(2)}(p=0)=m^2, \hspace{.4cm} \Gamma_A^{(2)}(p=\mu)=m^2+\mu^2,\nonumber\\
&\Gamma^{(2)}_{\bar c c}(p=\mu)=\mu^2.
\end{align}

We use the Taylor scheme for the coupling \cite{Taylor71}, defining
the coupling constant from the ghost-antighost-gluon vertex when the
ghost momentum is zero. Using the fact that this vertex has no quantum
corrections, one deduces that (\ref{no-renorm1}) is valid also for the
finite part of the renormalization factors.

\subsection{Strict perturbative results for 2-point correlation functions}

Let us first consider the calculation of the ghost self-energy in
arbitrary dimension using dimensional regularization.  Only one
diagram contributes (see second line of Fig.~\ref{diagrams}).  The
only difference with respect to standard Yang-Mills calculations is
the form of the bare gluon propagator (\ref{eq_propag_bare}).
Introducing Feynman parameters and performing internal momentum
integrals one easily arrives at the following expression for the ghost
2-point vertex:
\begin{align}
\label{gammacbare}
&\Gamma_{\bar c c}^{(2),\text{1 loop}}(p)=-\frac{g^2N
}{(4\pi)^{d/2}}\frac{p^2}{m^2}
\Gamma\Big(2-\frac{d}{2}\Big)\nonumber\\ &\times\int_0^1 dx \Bigg\{(x
m^2+x(1-x)p^2)^{d/2-2} \nonumber\\ &\times\Big(m^2+p^2(1-x)^2+\frac{x
  m^2+x(1-x)p^2}{2-d}\Big) \nonumber\\ &-(x(1-x)p^2)^{d/2-2} \Big(p^2
x^2+\frac{x(1-x)p^2}{2-d}\Big)\Bigg\}.
\end{align}
The remaining integral can be performed analytically in integer dimensions as discussed below.
Let us only mention here the UV divergent part when $d\to 4$:
\begin{equation}
\label{divghost}
\Gamma_{\bar c c}^{(2),\text{1 loop}}(p)\stackrel{d\to 4}{\sim}-\frac{3}{2} \frac{g^2N p^2}{16\pi^2 (4-d)}.
\end{equation}
This is in agreement with previous calculations in the literature
\cite{deBoer95,Gracey02}.

Let us now consider the 2-point vertex for gluons. The first diagram
contributing to this vertex function is the gluon tadpole (first
diagram in the first line of Fig.~\ref{diagrams}) and gives a
contribution:
\begin{equation}
\label{gammaa1}
\Gamma_{A,1}^{(2)}(p)=\frac{g^2 N}{(4\pi)^{d/2}} \frac{(d-1)^2}{d}
{m^{d-2}\Gamma(1-\frac{d}{2})}.
\end{equation}
The only effect of this diagram is to renormalize the mass.  The
second term contributing to the gluon self-energy is the ghost sunset
(second diagram in first line of Fig.~\ref{diagrams}) which reads:
\begin{equation}
\label{gammaa2}
\Gamma_{A,2}^{(2)}(p)=\frac{g^2N}{(4\pi)^{d/2}} p^{d-2}
\frac{\Big(\Gamma\Big(\frac{d}{2}\Big)\Big)^2
  \Gamma\Big(2-\frac{d}{2}\Big)}{(2-d)\Gamma(d)}.
\end{equation}
Finally, the third diagram contributing to the gluon self-energy is
the gluon sunset (third diagram in the first line of
Fig.~\ref{diagrams}). Again, introducing Feynman parameters, we obtain:
\begin{widetext}
\begin{align}
\label{gammaa3}
&\Gamma_{A,3}^{(2)}(p)=2 g^2 N \int_0^1 dx\int \frac{d^dq}{(2\pi)^d}\Bigg\{\frac{2 \left( x^2
   p^4 d+ \left(x^2+2\right)q^2 p^2+q^4\right)}{ m^2 d} \left(\frac{1}{\left(q^2+ x m^2+x (1-x) p^2 \right)^2}-\frac{1}{\left(m^2+q^2+x(1-x) p^2 \right)^2}\right)\nonumber\\
&-\frac{2 p^2+q^2}{\left(m^2+q^2+x(1-x) p^2 \right)^2}+\frac{
   (d+2)
   q^6+ \left(8 x^2-8 x+2 d \left(x^2-x+1\right)+5\right) q^4 p^2 +(d+2) 
   \left(x^2-x+1\right)^2 q^2 p^4}{d (d+2) m^4}\nonumber\\
&\times\left(\frac{1}{\left(
   m^2 (1-x)+q^2+x(1-x) p^2 \right)^2}-\frac{1}{\left(m^2+q^2+x(1-x)p^2 \right)^2}+\frac{1}{\left(x m^2+q^2+x(1-x)p^2 \right)^2}
-\frac{1}{\left(q^2+x(1-x) p^2\right)^2}\right)\Bigg\}.
\end{align}
\end{widetext}
The integral in the internal momentum $q$ can be done analytically in
arbitrary dimensions but the result is lengthy and not particularly
illuminating.  We can compute the divergent part of the gluon 2-point
function, which reads:
\begin{equation}
\Gamma_{A}^{(2),\text{1 loop}}(p)\stackrel{d\to 4}{\sim} \frac{g^2N }{16\pi^2 (4-d)}\left(\frac{3}{2}m^2-\frac{13}{3}p^2\right).
\end{equation}
Together with (\ref{divghost}), this leads to the determination of the
divergent part of the renormalization factors at one loop:
\begin{eqnarray}
 Z_{m^2}&=&1-\frac{35}{6} \frac{g^2 N}{16\pi^2} \frac{1}{4-d}\;,\nonumber\\
 Z_c&=&1+\frac 3 2 \frac{g^2 N}{16\pi^2} \frac{1}{4-d}\;,\nonumber\\
 Z_A&=&1+\frac{13}{3} \frac{g^2 N}{16\pi^2} \frac{1}{4-d}\;,
\end{eqnarray}
which coincide with previous results \cite{deBoer95,Gracey02}.

\subsection{Results in $d=4$}

When $d\to 4$ the expressions for correlation functions
(\ref{gammacbare},\ref{gammaa1},\ref{gammaa2},\ref{gammaa3}) diverge
in the UV.  In order to calculate the renormalized 2-point functions,
one must consider the sum of those self-energies with a bare
contribution with renormalization factors.  When $d$ approaches an
integer dimension, the remaining integrals can be performed
analytically.

Once the renormalization conditions are imposed (see
Eq.~\ref{rencond}) the renormalized functions $F(p)$ and $G(p)$ are
finite in the limit $d\to 4$ and read:
\begin{align}
\label{4dprops}
&G^{-1}(p)/m^2=s+1+ \frac{g^2 N s}{384 \pi ^2} \Big\{111s^{-1}-2 s^{-2}
\nonumber\\
&\hspace{.3cm}+(2-s^2)\log s+2(s^{-1}+1)^3\left(s^2-10 s+1\right) \log(1+s)\nonumber\\
&\hspace{.3cm}+(4 s^{-1} +1)^{3/2}
   \left(s^2-20 s+12\right)\log \left(\frac{\sqrt{4+s}-\sqrt{s}}{\sqrt{4
   +s}+\sqrt{s}}\right) \nonumber\\
&\hspace{1.3cm}-(s\to \mu^2/m^2)\Big\}, \nonumber \\
&F^{-1}(p)=1+\frac{g^2 N}{64 \pi ^2} \Big\{-s \log s +(s+1)^3 s^{-2} \log(s+1)\nonumber\\
&\hspace{1.3cm} - s^{-1}-(s\to \mu^2/m^2)\Big\},
\end{align}
where $s=p^2/m^2$.

\begin{figure}[tbp]
\epsfxsize=7.6cm
\epsfbox{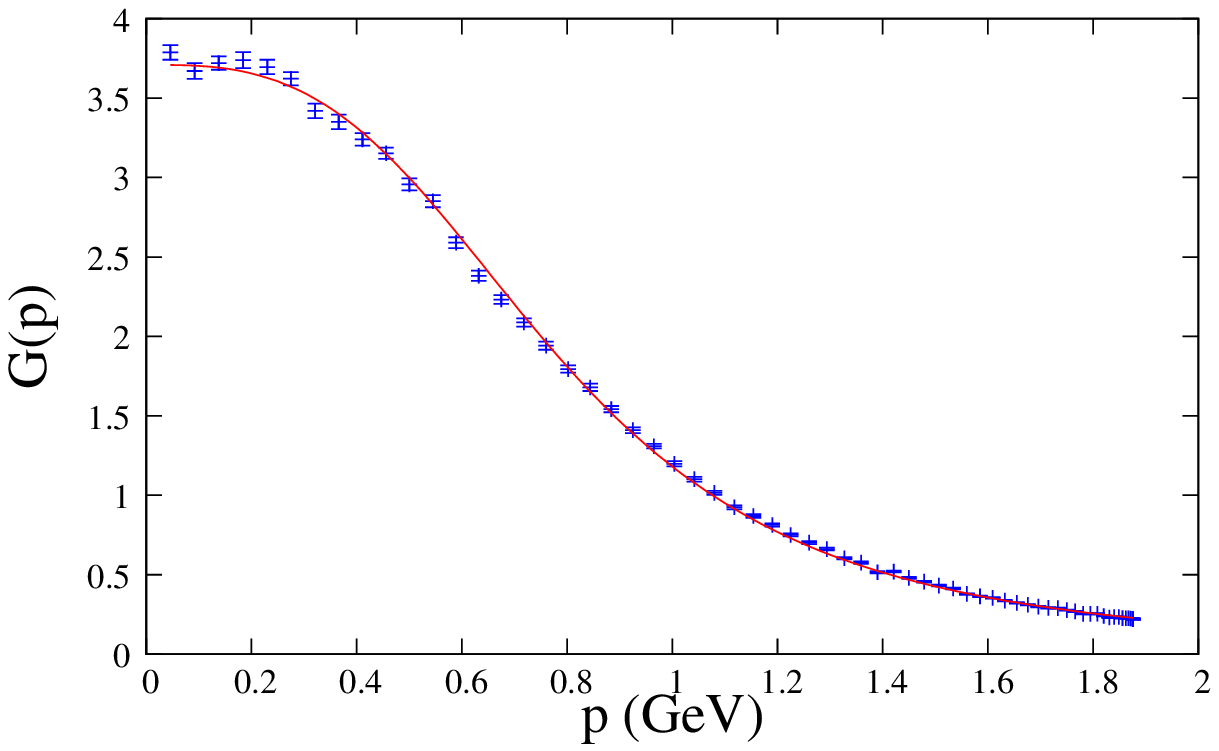}
\epsfxsize=7.6cm
\epsfbox{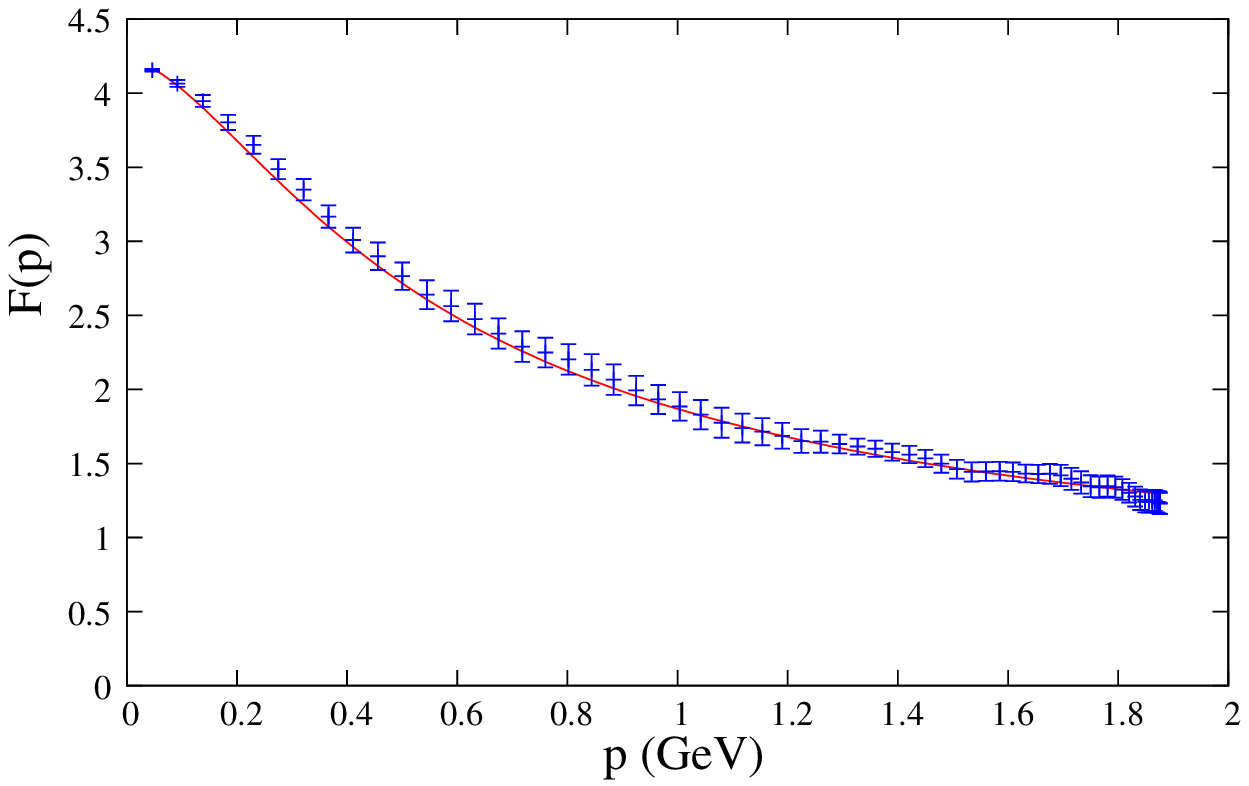}
\caption{\label{fig4dsu2}Four-dimensional correlation functions for
  $SU(2)$ gauge group. The results of the present work (red curve) are
  compared with lattice data of~\cite{Cucchieri_08b} (blue
  points).  Top figure: gluon propagator. Bottom figure: ghost
  dressing function.}
\end{figure}

In Fig.~\ref{fig4dsu2}, we compare these expressions for the $SU(2)$
gauge group with the lattice simulations of~\cite{Cucchieri_08b}. The best
choice of parameter is $g=7.5$ and $m=0.68$~GeV when the normalization
prescriptions are imposed at $\mu=1$~GeV. One observes that both gluon
and ghost propagators can be fitted with the same choice of parameters
in a very satisfactory way. Note that the normalization conditions
used in lattice simulations are not (\ref{rencond}). Accordingly, it
is necessary to introduce a global multiplicative renormalization
factor when comparing the curves.
\begin{figure}[tbp]
\epsfxsize=7.6cm
\epsfbox{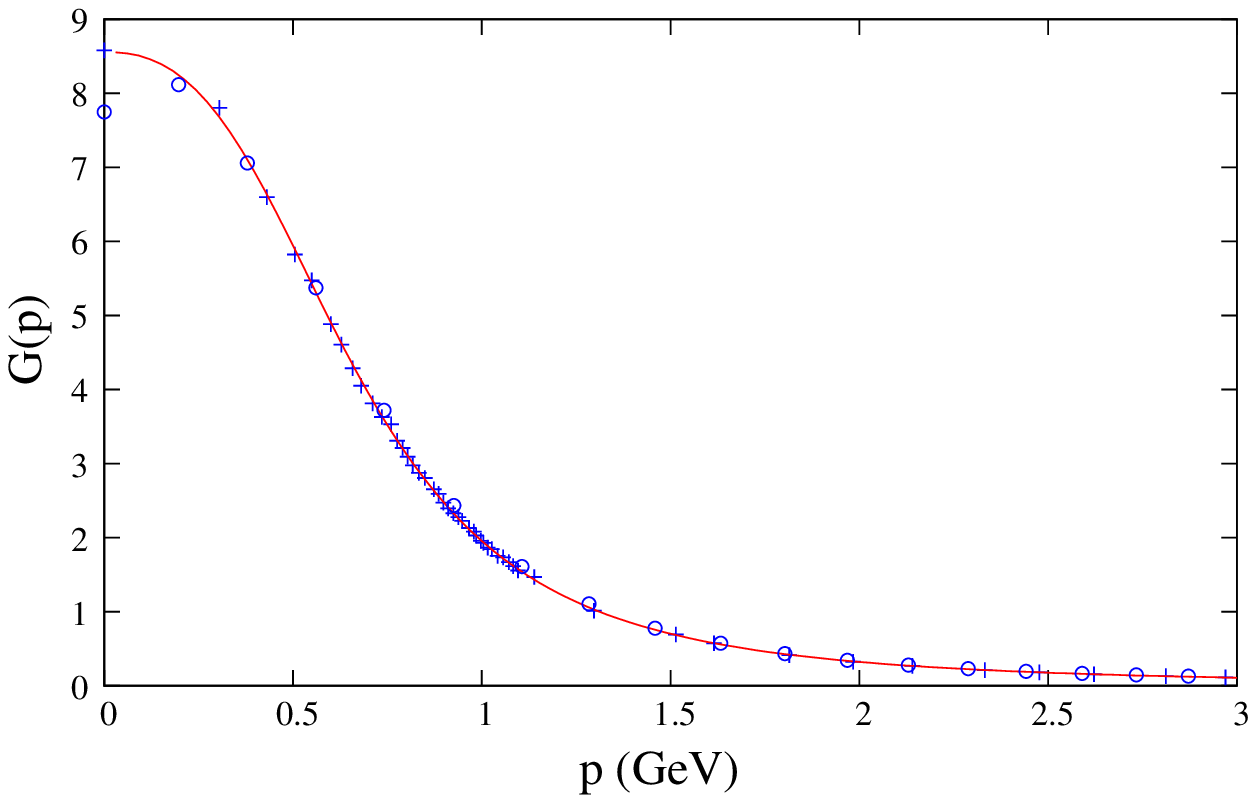}
\epsfxsize=7.6cm
\epsfbox{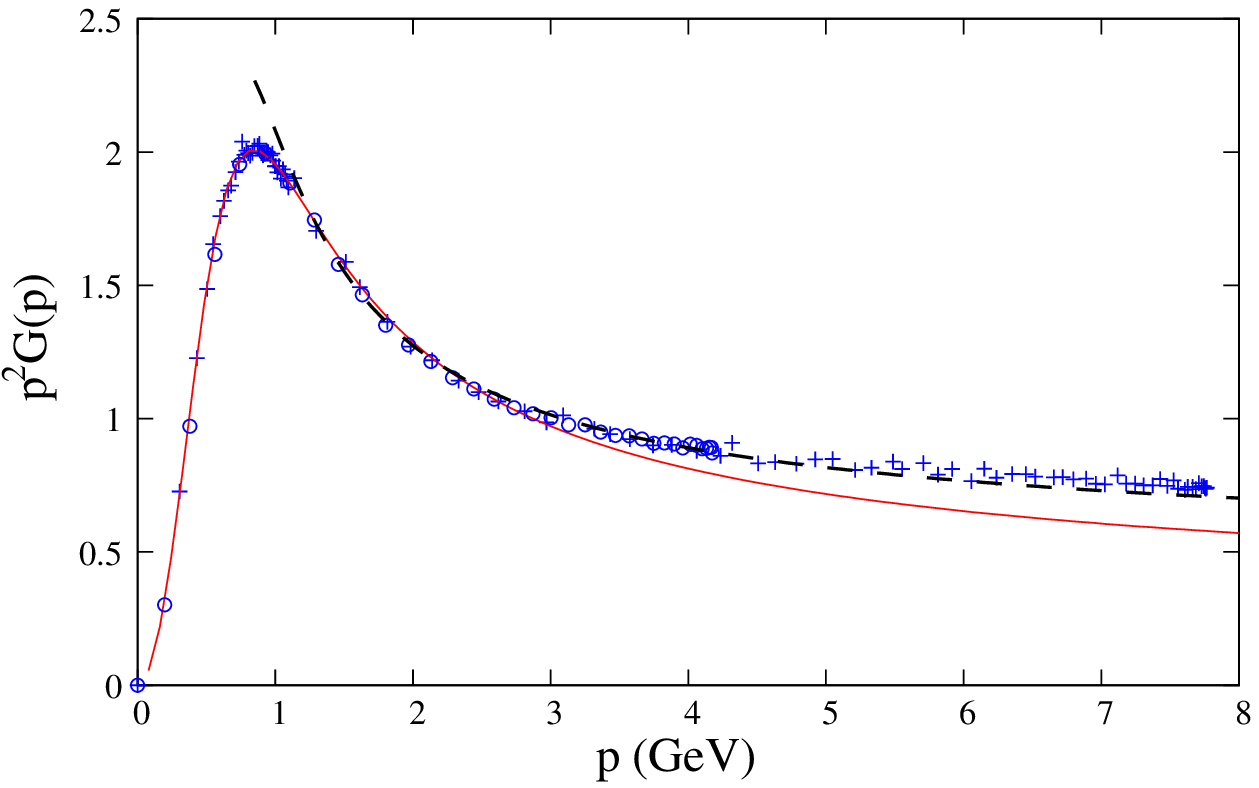}
\epsfxsize=7.6cm
\epsfbox{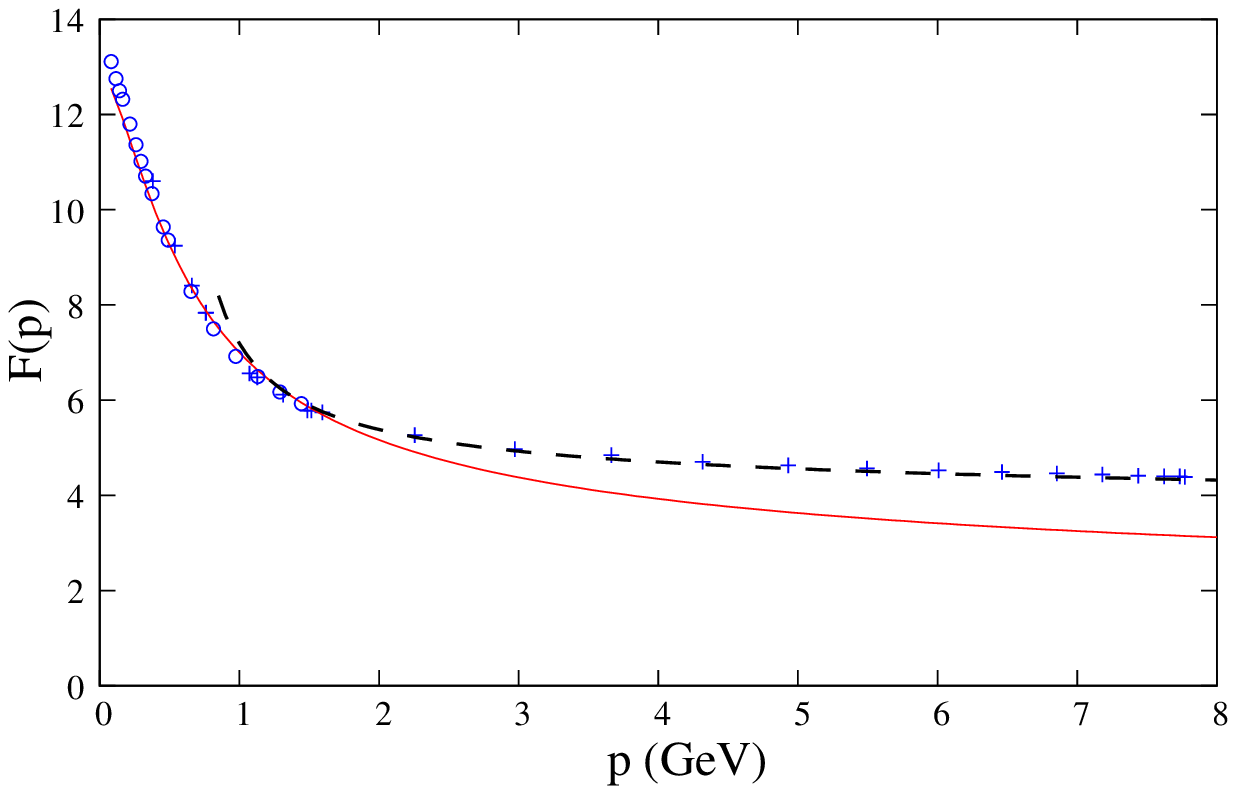}
\caption{\label{fig4dsu3}Four-dimensional correlation functions for
  $SU(3)$ gauge group. The results of strict perturbation theory (red
  solid curve) are compared with lattice data of~\cite{Bogolubsky09}
  (blue open circles) and~\cite{Dudal10} (blue crosses). The (black
  dashed) curve is obtained by the zero momentum prescription RG (see
  Sect.~\ref{zeromomscheme}). Top figure: gluon propagator. Middle
  figure: gluon propagator times $p^2$.  Bottom figure: ghost dressing
  function.}
\end{figure}

We have also compared our results with the data of two different
lattice studies~\cite{Bogolubsky09,Dudal10} for the $SU(3)$
group. These two data sets have different overall momentum scale and
we have rescaled the momenta of the data of~\cite{Bogolubsky09} for
superimposing them with those of~\cite{Dudal10}. Contrarily to the
$SU(2)$ case where data are only available in a small momentum
interval up to 1.9 GeV, the data for $SU(3)$ were computed for momenta
up to 8 GeV. Consequently one can explore the cross-over from standard
perturbative results when $p$ is significantly larger than $m$
(requiring RG methods) and an IR regime similar to the one analyzed in
the $SU(2)$ case. We represent in Fig.~\ref{fig4dsu3} the gluon
propagator and the ghost dressing function.  We present also the gluon
dressing function $p^2\,G(p)$ in order to make visible the UV
regime. The best choice of parameters is $g=4.9$ and $m=0.54$~GeV
(again with the renormalization prescription imposed at $\mu=1$ GeV)
and it leads to a very satisfying agreement for momenta $p \lesssim
2$~GeV. Beyond 2~GeV, the agreement is not as good but this is not a
surprise because expressions (\ref{4dprops}) are 1-loop results
obtained from a fixed coupling constant calculation in a fixed
renormalization point. It is well-known that in order to analyze the
regime $p \gg m$, one must take into account RG effects and in
particular the running of the coupling. The corresponding procedure is
presented in detail in the next sections.  Once RG effects are
included, the agreement is essentially within error bars for $p>m$ as
is also shown in Fig.~\ref{fig4dsu3}. In any case, it is obvious that
when $p\gg m$, the model (\ref{eq_lagrang}) reproduces correctly the
high momentum regime once RG effects are taken into account because it
just coincides in that case with the standard FP model.

A natural question to raise at this point is whether the parameters
found by the previous fitting procedure are compatible with other
determinations. We choose to compare $g$ with the results of
\cite{Boucaud08b} which make use of a very similar renormalization
scheme: the coupling is defined through the Taylor scheme and the
ghost renormalization factor is fixed by the second line of
Eq.~(\ref{rencond}). The only difference is that the gluon
renormalization factor is fixed by the condition 
$\Gamma_A^{(2)}(p=\mu)=\mu^2$. Because of the non-renormalization
theorem of the coupling, this amounts to a relation between the
coupling constant $g^{(\text l)}$ of  \cite{Boucaud08b} and our which
reads:
\begin{equation}
\label{eq_change_coupling}
g^{(\text l)}=\frac g{\sqrt{1+\frac{m^2}{\mu^2}}}.
\end{equation}
We extract from Fig. 1 of Ref. \cite{Boucaud08b} the value $g^{(\text
  l)}\simeq 3.5$ at $\mu=$1 GeV while the RHS of
(\ref{eq_change_coupling}) leads to 4.3. The error is about 20\%
which, as discussed below is the typical estimate of higher loop
corrections, see Sect.\ref{IR_suppress}.

An interesting feature of the 1-loop gluon propagator is that it is
increasing in the IR.  In fact, the inverse propagator behaves at
small momenta as $m^2+N
g^2p^2/(192\pi^2)\log(p^2/m^2)+\mathcal{O}(p^2)$.  This prediction of
our calculation has a very small effect for $d=4$ and it is not
visible in Figs.~\ref{fig4dsu2},\ref{fig4dsu3} but appears clearly in
$d=3$, as shown below.

\subsection{Results in $d=3$}

Let us now consider the three-dimensional case. As in $d=4$, the 2-point vertex functions can be
obtained in an explicit form:
\begin{equation}
\begin{split}
\Gamma_{\bar c c}^{(2),\text{1\,loop}}(p)=\frac{g^2 m N}{32\pi\sqrt{s}}&\Big(\pi s^2 + 2\sqrt{s}(1 -  s) \\
&- 2 (s + 1)^2\arctan\left(\sqrt{s}\right)\Big),
\end{split}
\end{equation}
\begin{equation}
\begin{split}
&\Gamma_{A}^{(2),\text{1\,loop}}(p)=\frac{g^2 m N}{128\pi s^{3/2}}\Big(-4\left(5 s^2 + 7 s - 1\right)\sqrt{s} \\
&+\pi\left(s^2 - 2\right) s^2 -4 (s + 1)^2\left(s^2 - 6 s +
          1\right)\arctan\left(\sqrt{s}\right)  \\
&+2s (s + 4)\left(s^2 - 12 s + 8\right)\arctan\left(\frac{\sqrt{s}}{2}\right)   \Big),
\end{split}
\end{equation}
where again $s=p^2/m^2$.

These expressions are finite because in $d=4$ all divergences are
logarithmic and lowering the dimension reduces the level of UV
divergence of any diagram.  Accordingly, a possible renormalization
scheme could be to use as renormalized parameters just the bare ones
(that are finite). However it is well known that is is more convenient
to chose a renormalization scheme where renormalized parameters are
defined at a running scale. This corresponds to a finite
renormalization with respect to the bare parameters.  This is
convenient in order to avoid large perturbative corrections. Moreover,
choosing renormalized parameters at a running scale will allow us to
perform a RG analysis in the next sections (that improves considerably
the results).  In the present section we use the renormalization
conditions~(\ref{rencond}).

\begin{figure}[tbp]
\epsfxsize=7.6cm
\epsfbox{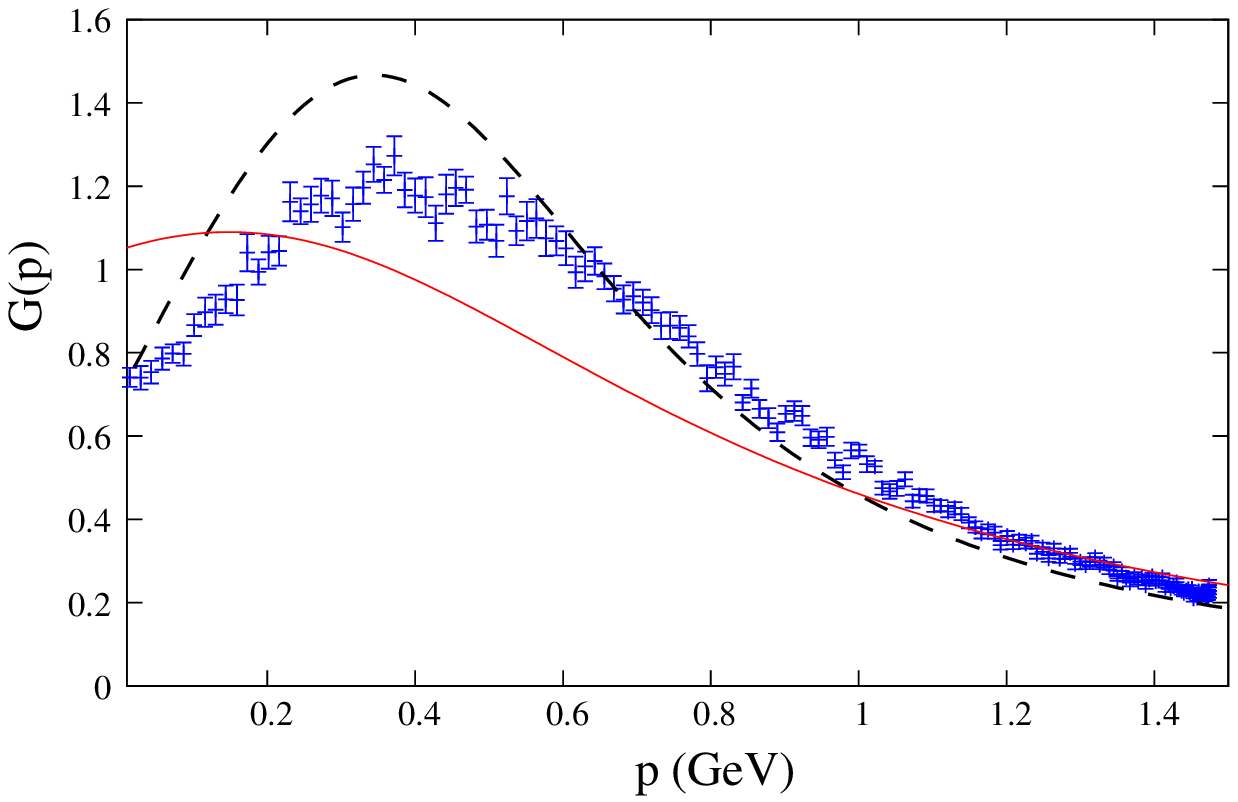}
\epsfxsize=7.6cm
\epsfbox{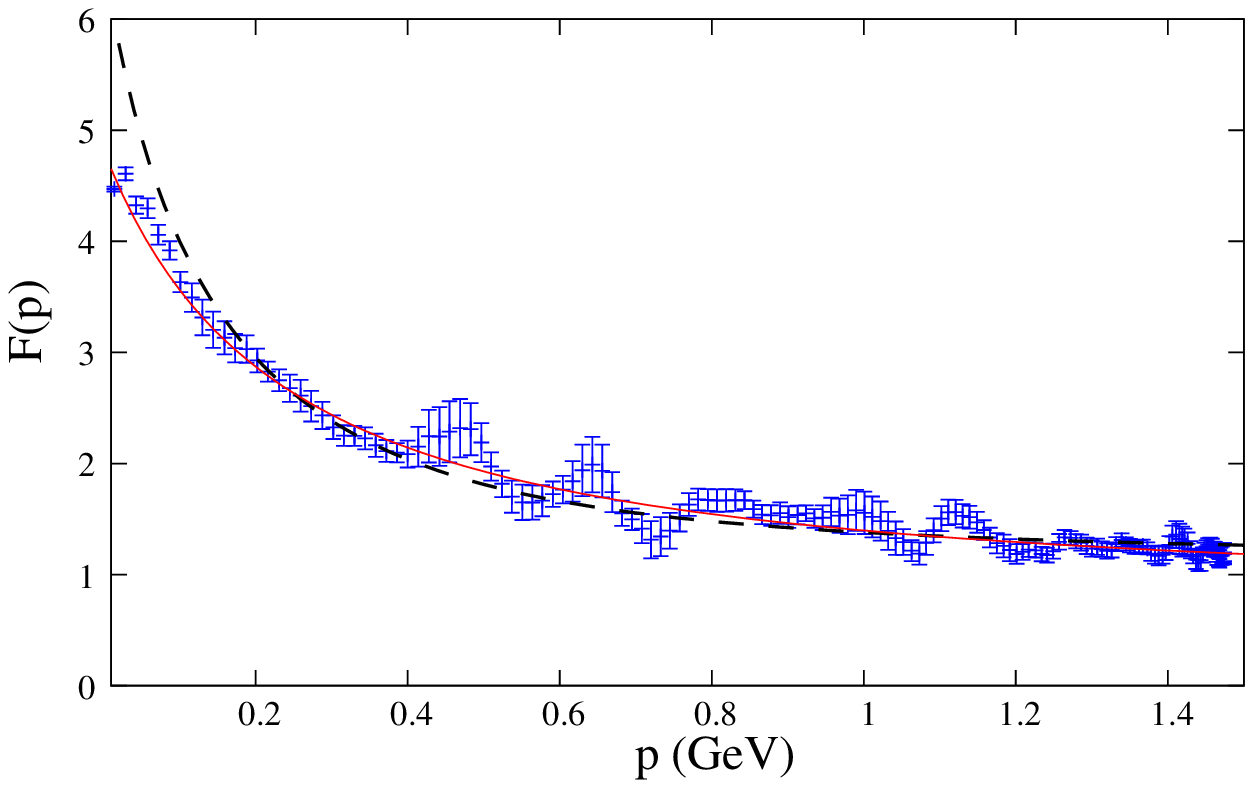}
\caption{\label{fig3d}Three-dimensional correlation functions for
  $SU(2)$ gauge group. The results of strict perturbation theory (red
  solid curve) are compared with lattice data of~\cite{Cucchieri_08b}
  (blue points). The (black dashed) curve is the IR-safe RG improved
  result (see Sect. \ref{irsafescheme}).  Top figure: gluon
  propagator. Bottom figure: ghost dressing function.}
\end{figure}

The model (\ref{eq_lagrang}) is able to account for the main features
of gluon and ghost propagators found in lattice simulations. In
Fig.~\ref{fig3d} the results of our calculation with the best fit
parameters $g=3.7$~$\sqrt{\text{GeV}}$ and $m=0.89$~GeV for
$\mu=1$~GeV are compared with $d=3$ simulations performed with the
gauge group $SU(2)$~\cite{Cucchieri_08b}. We observe that the best fit for
gluon and ghost propagators are not as good as in $d=4$.  We will show
below that the inclusion of RG effects improve significantly one-loop
results.
 In any case, our calculation reproduces the finite IR gluon
 propagator and ghost dressing function. It also reproduces the increasing behavior of the gluon propagator in the IR. Indeed, an
 expansion of the inverse propagator at low momentum leads to
 $m^2-Ng^2p/64+\mathcal{O}(p^2)$.

\section{Renormalization Group analysis}

We showed in the previous section that the 1-loop perturbative results
compare very well with the lattice data up to energies of order 2 GeV
and get worse at higher energies. This is not a surprise since in
$d=4$ the perturbation theory generates logarithms of the momentum
divided by some momentum scale (the mass or the renormalization point
$\mu$). The pure perturbation theory fails at energies higher
than 2 GeV because these logarithms are large and one needs to
implement the ideas of RG. Actually since we have massless modes
(ghosts) in the theory, it is expected a priori that the RG is
necessary also for $p \ll m$. We will see however that in this
particular model, the IR behavior is milder than expected, essentially
because ghosts interact by the exchanging gluons which are massive.

For $d<4$, the UV behavior does not present large logarithms, and
consequently the use of the RG is not mandatory in that
regime. However, in practice, taking into account RG effects may
improve the quantitative results. This idea is natural, but we are not
able to test it because there is no UV data available in $d=3$ or
$d=2$.  For $d<4$, a RG adapted for the IR is much more important than
for $d=4$ and at the end of the present section we show that such a RG
procedure improves considerably the results in $d=3$. The comprehension
of the $d=2$ case also improves considerably by the inclusion of RG
effects but this discussion is postponed to Sect.~\ref{2dcase}.

\subsection{Vanishing-momentum prescription scheme in $d=4$}
\label{zeromomscheme}
We recall the main steps of renormalization, mainly to fix the
notations. We concentrate in this section on the renormalization
prescriptions that were described in Sect.~\ref{themodel}. We define the
$\beta$ functions and anomalous dimensions as:
\begin{align}
\beta_g(g,m^2)&=\mu\frac{dg}{d\mu}\Big|_{g_B, m^2_B},\\
\beta_{m^2}(g,m^2)&=\mu\frac{dm^2}{d\mu}\Big|_{g_B, m^2_B},\\
\gamma_A(g,m^2)&=\mu\frac{d\log Z_A}{d\mu}\Big|_{g_B, m^2_B},\\
\gamma_c(g,m^2)&=\mu\frac{d\log Z_c}{d\mu}\Big|_{g_B, m^2_B}.
\end{align}
We can then use the RG equation:
\begin{equation}
\begin{split}
\Big( \mu \partial_\mu -\frac 1 2 &(n_A \gamma_A+n_c \gamma_c)\\&+\beta_g \partial_{g}+
\beta_{m^2}\partial_{m^2}\Big)\Gamma^{(n_A,n_c)}=0,
\end{split}
\end{equation}
to relate the 2-point
vertex functions at different scales:
\begin{align}
\label{eq_cal_AA}\Gamma_{A}^{(2)}(p,\mu,&g(\mu),m^2(\mu))\nonumber\\
&=z_A(\mu)\Gamma_{AA}^{(2)}(p,\mu_0,g(\mu_0),m^2(\mu_0)),\\
\label{eq_cal_cc}\Gamma_{\bar c c}^{(2)}(p,\mu,&g(\mu),m^2(\mu))\nonumber\\
&=z_c(\mu)\Gamma_{\bar
  c c}^{(2)}(p,\mu_0,g(\mu_0),m^2(\mu_0)).
\end{align}
where $g(\mu)$ and $m^2(\mu)$ are obtained by integration of the
beta functions with initial conditions given at some scale $\mu_0$ and:  
\begin{equation}
\label{eq_def_z_phi}
\begin{split}
\log z_A(\mu)&=\int_{\mu_0}^\mu\frac
     {d\mu'}{\mu'}\gamma_A\left(g(\mu'),m^2(\mu')\right),\\ \log
     z_c(\mu)&=\int_{\mu_0}^\mu\frac
     {d\mu'}{\mu'}\gamma_c\left(g(\mu'),m^2(\mu')\right).
\end{split}
\end{equation}
Using the normalization conditions described in Sect.~\ref{themodel}, we then get:
\begin{equation}
\Gamma_{A}^{(2)}(p,\mu_0,g,m^2)=\frac{p^2+m^2(p)}{z_A(p)},
\end{equation}
\begin{equation}
\Gamma_{\bar c c}^{(2)}(p,\mu_0,g,m^2)=\frac{p^2}{z_c(p)}.
\end{equation}

The renormalization scheme considered here leads to a relation between
the $\beta$ functions and the anomalous dimensions $\gamma_A$ and
$\gamma_c$.  As explained before, in the Taylor scheme the
non-renormalization theorem (\ref{no-renorm1}) is also valid for the
finite parts. This leads to the following expression for the $\beta$
function of the coupling constant:
\begin{equation}
\label{eq_def_beta_g}
\beta_g(g,m^2)= g \left(\frac{\gamma_A(g,m^2)}2+\gamma_c(g,m^2)\right).
\end{equation}
The choice of a normalization prescription at vanishing momentum for
the gluon 2-point vertex simplifies greatly the analytical treatment
because as shown in the Appendix \ref{sect_non-renorm} the following
relation is satisfied for the finite parts of the propagators and
dressing functions:
\begin{equation}
\Gamma^{(2)}_{A}(p=0)=Z_Am_B^2F_B(p=0).
\end{equation}
It is then easy to prove that:
\begin{equation}
\label{eq_def_beta_m2}
\beta_{m^2}(g,m^2)=m^2\gamma_A(g,m^2).
\end{equation}

The perturbative results together with the renormalization scheme
(\ref{rencond}) enables us to compute the anomalous dimensions. We get:
\begin{equation}
\label{eq_gamma_A}
\begin{split}
\gamma_A=&-\frac{g^2 N}{192 \pi ^2 t^3} \Big(t (34 t^2-175 t +6)-2  t^5 \log t\\
&+2 (t+1)^2 (2 t^3-11 t^2+20 t-3) \log (t+1)\\
&+2 t^{3/2} \sqrt{t+4} \left(t^3-9 t^2+20
   t-36\right) \\
 &\times\log
   \Big(\frac{\sqrt{t+4}-\sqrt{t}}{\sqrt{t+4}+\sqrt{t}
   }\Big)\Big),
\end{split}
\end{equation}
\begin{equation}
\label{eq_gamma_c}
\begin{split}
\gamma_c=-\frac{g^2 N}{32 \pi ^2 t^2} \Big(&2(t+1)t -t^3 \log t\\
&+(t+1)^2 (t-2) \log (t+1)\Big),
\end{split}
\end{equation}
where $t=\mu^2/m^2$.

These expressions for the anomalous dimensions lead, together with
Eqs.(\ref{eq_def_beta_g}) and (\ref{eq_def_beta_m2}), to the
expressions of the $\beta$ functions. Observe that these $\beta$ functions 
are not only function of the coupling constant $g$ (as for example if 
we use the $\overline{MS}$ scheme), but also depend on the dimensionless ratio $\mu^2/m^2$. In principle, once these flow
equations are integrated with some appropriate initial conditions, we
would need to make another integral to determine $z_A(\mu)$ and
$z_c(\mu)$ (see eq. (\ref{eq_def_z_phi})) in order to determine
 the inverse propagators. However, since the flows of $g$ and
$m^2$ are expressed in terms of the anomalous dimensions for the
ghost and gluon fields, these integrals can be done
explicitly:
\begin{equation}
\label{eq_int_za}
z_A(\mu)=\frac{m^2(\mu)}{m^2(\mu_0)},
\end{equation}
\begin{equation}
\label{eq_int_zc}
z_c(\mu)=\frac{g(\mu)}{g(\mu_0)}\sqrt{\frac{m^2(\mu_0)}{m^2(\mu)}}.
\end{equation}

Studying the behavior of these anomalous dimensions for $\mu$ much larger and much smaller than $m$ together with eq. (\ref{eq_def_beta_g}), we find:
\begin{equation}
\beta_g\sim
\begin{cases}
-\frac{g^3N}{16\pi^2}\frac{11}3\qquad \text{if}\qquad \mu\gg m,\\
-\frac{g^3N}{16\pi^2}\frac1{12}\qquad \text{if}\qquad \mu\ll m.
\end{cases}
\end{equation}
In the UV ($\mu\gg m$) we obtain the standard, universal, $\beta$ function. This is in contrast to
other approaches where masses are introduced in Yang-Mills theory but modifying the UV behavior of the model \cite{Cornwall79}. 
In the IR ($\mu\ll m$), we observe that, although smaller in absolute value than in the UV,
$\beta_g$ remains negative which means that the coupling
constant diverges at a finite energy scale: we encounter a Landau pole
in the integration. 

However, we stress that there are no large quantum corrections when
$p\ll m$ (at least when $d>2$) because there is no IR divergences. In
this configuration, a natural scheme is to use strict perturbation
theory for $p \lesssim m$ and a RG improvement when $p \gtrsim m$.
The absence of IR divergences might be surprising since the ghosts are
massless. A simple way of understanding this is to observe that all
diagrams, except those with a single ghost loop, include at least one
gluon propagator, which regularize the IR behavior (this issue is
discussed in more detail in Appendix \ref{appendix_suppress}). In this
sense a purely perturbative calculation for momenta $p\lesssim$ 1 GeV
and a RG improvement for UV regime ($p \gtrsim$ 1 GeV) is fully
justified. In practice, since the RG flow presents a Landau pole at an
energy scale not far from 1GeV, we want to make the matching at a
higher energy, but lower than 2 GeV where strict perturbation theory
begins to become untrustable. A compromise is to perform the matching
condition at 1.5 GeV and this gives impressively good results, as can
be seen in Fig.~\ref{fig4dsu3}.

\subsection{IR safe scheme in $d=4$}
\label{irsafescheme}

A drawback of the procedure described in the previous section is that there is some arbitrariness on the choice of the matching scale
to join strict perturbative and RG results. In this sense a unified procedure, valid at all scales would be more satisfactory
and this motivates the search for an IR safe RG scheme.

In this respect, it is interesting to note that the IR Landau pole
that appears in the previous RG scheme is not a physical effect but is
actually a consequence of our renormalization scheme
(\ref{rencond}). Indeed, the prescriptions we chose there are {\it
  not} consistent, if $\mu$ is too small, with the 1-loop results (and
with the actual behavior observed in lattice simulations in $d=3$)
that lead to an increasing propagator in the IR (see
Sect.~\ref{strictperturb}). In order to avoid this problem and be able
to take renormalization prescriptions at arbitrary scales, we present
here a scheme which is not based on an explicit constraint of the
2-points vertex function but which comes from the nonrenormalization
theorem (\ref{no-renorm2}).  In practice, we replace the first
condition in (\ref{rencond}) by imposing the relation:
\begin{equation}
\label{newrenormcond}
Z_AZ_cZ_{m^2}=1,
\end{equation}
not only for the divergent parts but also for the finite parts.  The
renormalized mass $m^2$ does not correspond any more to the value of
the renormalized propagator at zero momentum and accordingly there is
no contradiction with the increase of the propagator in
the IR.

Within this new scheme, few formulas must be modified. The $\beta$
function for the mass is now changed to:
\begin{equation}
\label{m2IRscheme}
\beta_{m^2}(g,m^2)=m^2(\gamma_A(g,m^2)+\gamma_c(g,m^2)).
\end{equation}
The anomalous dimension for the
ghost is unchanged (see eq. (\ref{eq_gamma_c})) while the anomalous
dimension for the gluon field reads:
\begin{equation}
\begin{split}
\gamma_A=&\frac{g^2 N}{96 \pi ^2 t^3}  \Big(-(t-2)^2 (2 t-3)
  (t+1)^2 \log (t+1) \\
&+(-17 t^2+74
   t-12)t+t^5 \log (t)\\
&-t^{3/2} \sqrt{t+4} \left(t^3-9
   t^2+20 t-36\right) \\
&\times\log
   \Big(\frac{\sqrt{t+4}-\sqrt{t}}{\sqrt{t+4}+\sqrt{t}
   }\Big)\Big),
\end{split}
\end{equation}
where, as before, $t= \mu^2/m^2$.
Finally, the expressions relating $z_A(\mu)$ and $z_c(\mu)$ to the
coupling constants read now:
\begin{equation}
\label{eq_int_za_IRsafe}
z_A(\mu)=\frac{m^4(\mu)}{m^4(\mu_0)}\frac{g^2(\mu_0)}{g^2(\mu)},
\end{equation}
\begin{equation}
\label{eq_int_zc_IRsafe}
z_c(\mu)=\frac{g^2(\mu)}{g^2(\mu_0)}\frac{m^2(\mu_0)}{m^2(\mu)}.
\end{equation}

The UV universal behavior of the $\beta$ function (for $\mu\gg m$) is
unchanged as expected.  However, we observe that the beta function for
the coupling constant in the IR is now {\em positive}:
\begin{equation}
\label{irbeta}
\beta_g\sim \frac{g^3N}{16\pi^2}\frac 1 6\qquad \text{if}\qquad\mu\ll m.
\end{equation}
We need to insist on the fact that the universality of the beta
function is only valid for mass-independent schemes or for $\mu \gg m$
up to two-loop order. In particular, for $\mu \lesssim m$ there is no
reason for the beta function to be scheme independent and it is
typically not the case.  The result (\ref{irbeta}) means that in the
present scheme, the IR behavior is possibly safe. Actually, depending
on the initial conditions of the flow we do, or do not, hit a Landau
pole. (Obviously, if we start with a vanishing mass, we retrieve the
standard 1-loop calculation and we do hit the Landau pole, so a
sufficiently large initial mass is necessary to be IR safe).
\begin{figure}[tbp]
\epsfxsize=7.6cm
\epsfbox{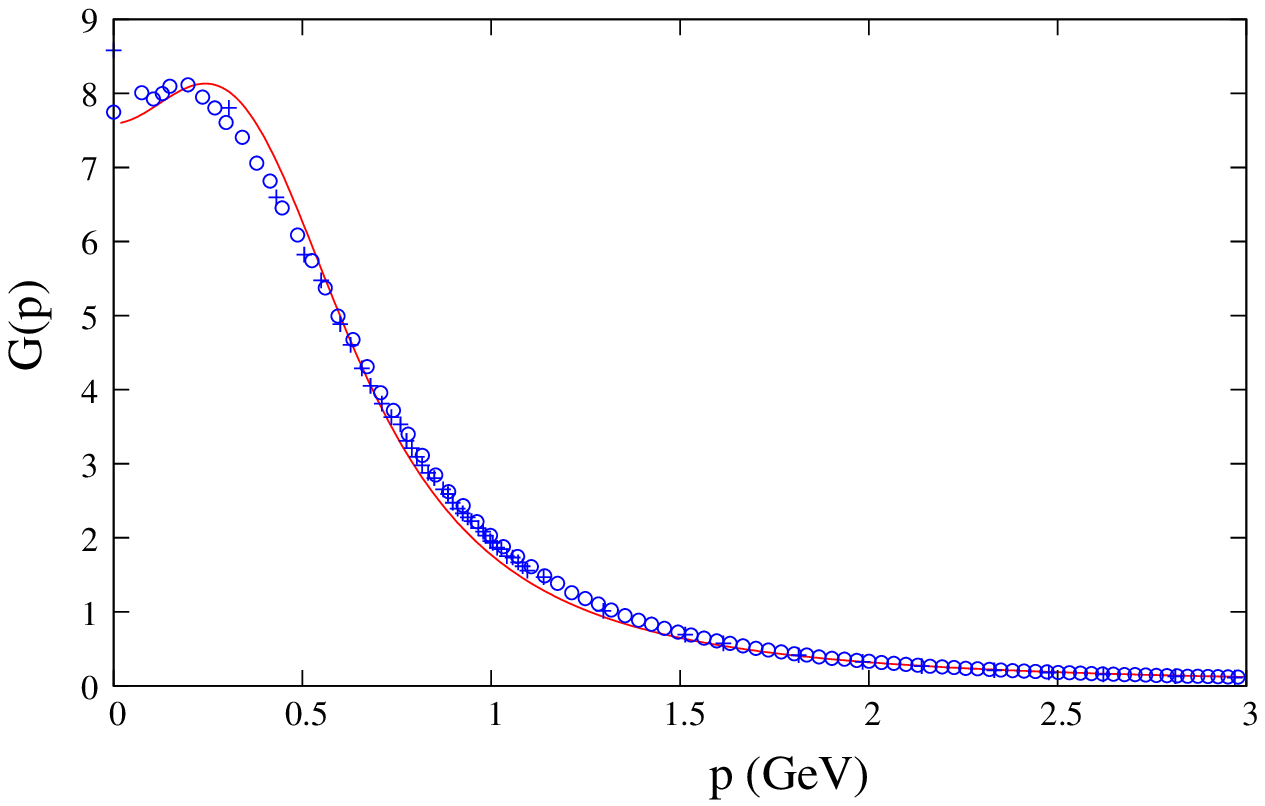}
\epsfxsize=7.6cm
\epsfbox{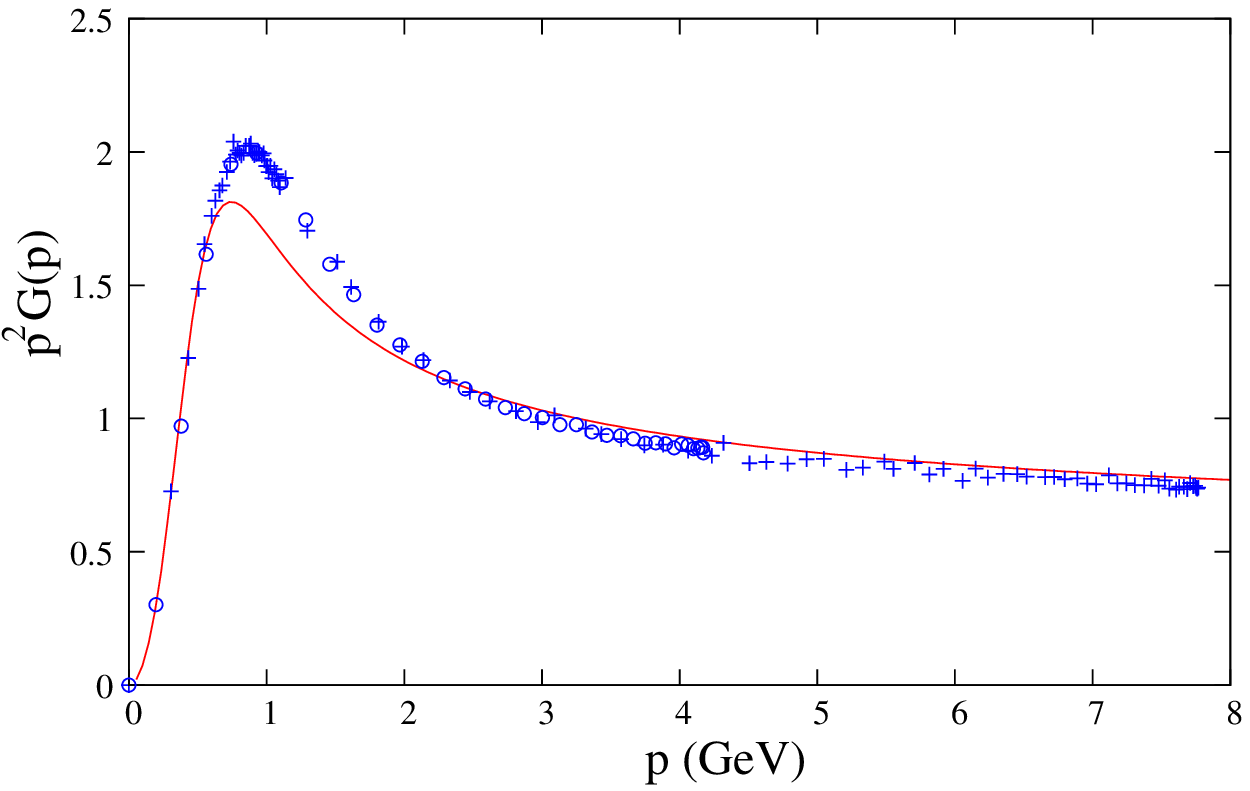}
\epsfxsize=7.6cm
\epsfbox{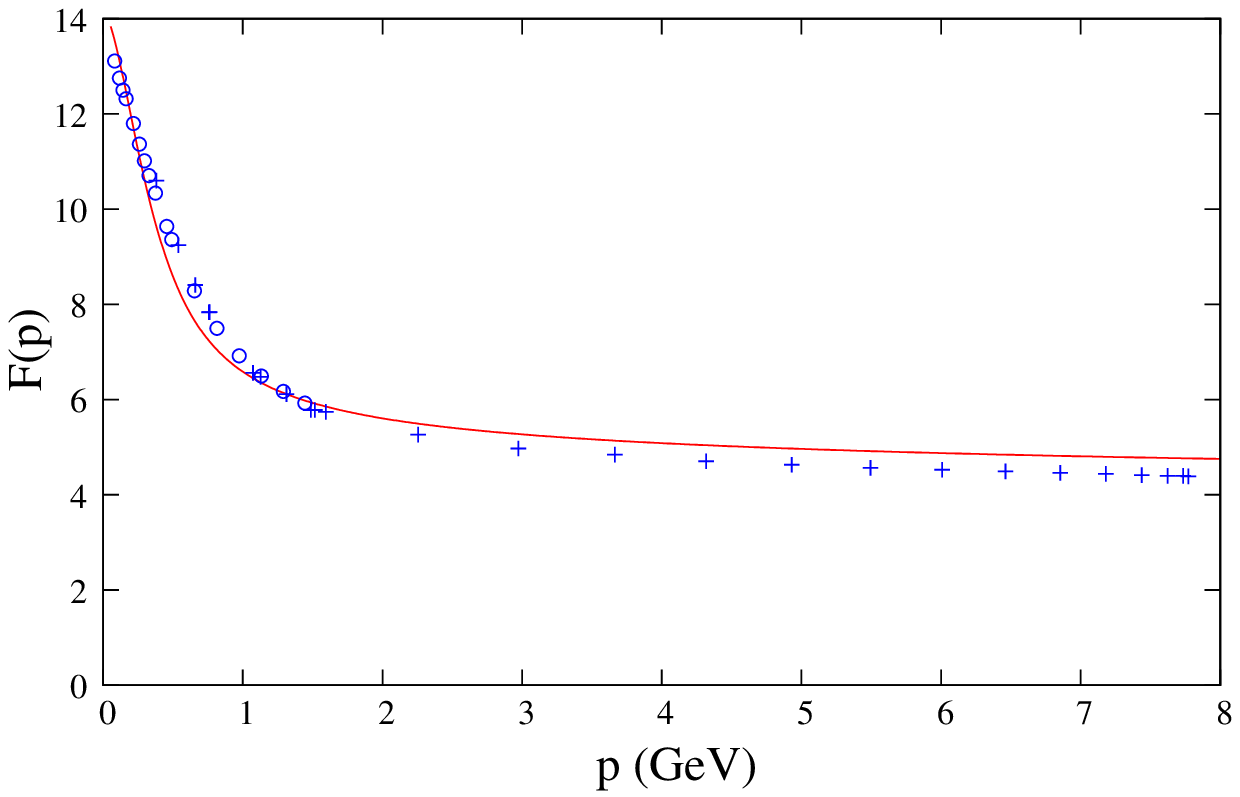}
\caption{\label{fig4drg}Four-dimensional correlation functions for
  $SU(3)$ gauge group. The results obtained by the IR safe RG (see
  Subsec.~\ref{irsafescheme}) (red solid line) are compared with
  lattice data of~\cite{Bogolubsky09} (blue open circles)
  and~\cite{Dudal10} (blue crosses).  Top figure: gluon
  propagator. Middle figure: gluon propagator times $p^2$.  Bottom
  figure: ghost dressing function.}
\end{figure}

A second consequence of this IR behavior is that if the flow does not
hit a Landau pole, then the coupling constant is attracted towards
zero in the IR: the massive gaussian fixed point is attractive. This
is a very appealing property because this justifies the use of
perturbation theory in the regime $p \ll m$, although its use at
intermediate momenta ($p\sim m$) is more delicate. The validity of
perturbation theory at intermediate momenta is discussed in
Sect.~\ref{IR_suppress}.

In Fig.~\ref{fig4drg} the results for the correlation functions
obtained from the numerical integration of the RG equations for the
coupling and mass is presented. The best choice of parameters at the
scale $\mu=$~1 GeV is $g=$~3.7 and $m=$~0.39 GeV.  One observes a very
good agreement of the 1-loop RG result when compared with lattice
simulations, particularly, as expected, in the UV and in the IR. The
only region where a significative departure is seen is for
intermediate momenta (1 GeV $\lesssim p\lesssim$ 2 GeV) for the gluon
dressing function. The results, however, are not as good as those
obtained with the scheme of Sect.~\ref{zeromomscheme}. This may be
surprising and we now explain why it is not so. First, as already
mentioned and explained in detail in the Appendix
\ref{appendix_suppress}, there are no IR divergences and consequently
the use of RG is not mandatory for $p\lesssim m$. Second, the results
of the previous section have been obtained by a mixed treatment, by
using the best fit of strict perturbation theory and imposing a
matching with RG results in the UV.  The results discussed here are
obtained with a unified treatment, valid at all scales.  This is a
more challenging situation because we do not have the freedom of
choosing the matching scale.  In this sense the IR safe scheme gives a
more satisfying construction. Third, as discussed in
Sect.~\ref{IR_suppress} even if the relevant coupling in the present
model is never large, it is nevertheless not very small. Accordingly
the result obtained in the present section are quite satisfactory and
the surprise is more the level of precision achieved in the previous
one. A possible explanation is that the choice of RG prescriptions
that can be directly read from correlation functions (and not from an
implicit RG scheme as done in (\ref{newrenormcond})), may in practice
reduce the contributions of higher loop corrections. In any case we
may expect that two-loop corrections should improve considerably the
present results and correspondingly reduce the scheme dependence of
them. Finally, the coupling constant compares much better with the one
extracted from lattice simulations (see
Eq.~(\ref{eq_change_coupling})). Taking into acount the difference of
renormalization scheme, we now find, at $\mu=1$ GeV, $g^{(\text l)}=3.4$
which compares very well with the lattice data 3.5 \cite{Boucaud08b},
see Eq.(\ref{eq_change_coupling}).

\subsection{The IR behavior in arbitrary dimensions}
\label{sect_ir}
Before discussing the results in $d=3$, it is convenient to study the
IR ($\mu \ll m$) behavior of RG functions of the model in arbitrary
dimensions. We saw in the previous subsection, that in that regime,
the beta function for the coupling constant is positive and
accordingly the gaussian IR fixed point is attractive. When $d\neq 4$,
the coupling constant $g$ has dimension $[\text{energy}]^{(4-d)/2}$
and it is convenient to introduce the dimensionless coupling
\begin{equation}
\label{gtilde}
 \tilde g(\mu)= g(\mu) \mu^{(d-4)/2},
\end{equation}
in order to analyze a process with characteristic momentum $\mu$.  Of
course, it may be appropriate to use powers of the mass in order to
define a dimensionless coupling.  This point is discussed below (see
Sect.~\ref{IR_suppress}), but for the moment let us concentrate on the
choice~(\ref{gtilde}) which is the natural definition of the
dimensionless coupling in the UV regime ($\mu\gg m$).  The
$\beta$-function associated with $\tilde g$ is:
\begin{equation}
 \beta_{\tilde g} = \mu \frac{\partial(g  \mu^{(d-4)/2})}{\partial \mu}=-\frac{4-d}{2} \tilde g+ \mu^{(d-4)/2}\beta_g.
\end{equation}
For dimensional reasons, the function $\beta_{\tilde g}$ is only a
function of $\tilde g$ and of the dimensionless squared mass:
\begin{equation}
 \tilde m^2= m^2 \mu^{-2}.
\end{equation}
For any $d<4$, as well known, in the UV regime ($\mu \gg m$ or,
equivalently $\tilde m^2 \ll 1$), the coupling $\tilde g$ approaches
rapidly a gaussian fixed point because of the dominance of the
dimensional term $-\frac{4-d}{2} \tilde g$. The case $d=4$ is also
dominated by a gaussian fixed point in the UV, because of the negative
sign of the $g^3$ term. At intermediate momenta $\mu \sim m$, the
expressions of the $\beta_{\tilde g}$ function is very complicated in
arbitrary dimensions and its (non illuminating) expression involves
special functions. However in the IR regime $\mu \ll m$ things become
easier again, and one can give an explicit expression for the
$\beta_{\tilde g}$ function at one-loop:
\begin{equation}
\label{IRbetaanyd}
\beta_{\tilde g}=-\frac{4-d}{2} \tilde g+\frac{ (4-d) (d-2) \pi
  ^{\frac{3-d}{2}}}{2^{2 d+1}\Gamma \left(\frac{d+1}{2}\right)\sin \big(\frac{(4-d) \pi
  }{2}\big)}N\,\tilde g^3.
\end{equation}
The coefficient of the $\tilde g^3$ term is regular (and non zero) in all dimensions $d>0$. One observes the appearence of
an IR fixed point $\tilde g$:
\begin{equation}
 N\,\tilde g^2_*=\frac{4^{d}\Gamma \left(\frac{d+1}{2}\right)\sin \left(\frac{(4-d) \pi
  }{2}\right)}{(d-2) \pi^{\frac{3-d}{2}}}.
\end{equation}
When $d\to 4$, the expression (\ref{IRbetaanyd}) approaches the $d=4$
result (\ref{irbeta}) and the fixed point value approaches the gaussian
limit $\tilde g_*^2=0$.

The coherence of these results relies on the hypothesis that $\tilde m^2\to \infty$ when $\mu \to 0$. To check this hypothesis, we
write the $\beta$ function for the dimensionless mass $\tilde m^2$:
\begin{equation}
 \beta_{\tilde m^2} = \mu \frac{\partial(m^2 \mu^{-2})}{\partial
   \mu}=-2 \tilde m^2+ \mu^{-2}\beta_{m^2}.
\end{equation}
As before, the UV regime (where $\tilde g\to 0$) is compatible with $\tilde m^2 \to 0$ and this can already be seen in the $\sim-2 \tilde m^2$
term. The intermediate momentum regime ($\mu \sim m$) gives complicated expressions, but the IR regime again simplifies to:
\begin{equation}
\label{eq_m2IR}
 \beta_{\tilde m^2} = -2 \tilde m^2+N\,\tilde g^2 \tilde m^2 \frac{
   (4-d) (d-2) \pi ^{\frac{3-d}{2}}}{4^{d}\Gamma
   \left(\frac{d+1}{2}\right)\sin \big(\frac{(4-d) \pi }{2}\big)}.
\end{equation}
Note that the pre-factor of $\tilde g^2 \tilde m^2$ of this expression
is twice the pre-factor of $\tilde g^3$ in the eq.~(\ref{IRbetaanyd}).
Consequently when $\tilde g^2$ approaches its IR fixed point:
\begin{equation}
 \beta_{\tilde m^2} \sim (2-d) \tilde m^2.
\end{equation}
For $d>2$, this implies that a regime with $\tilde m^2\to \infty$ and
$\tilde g \sim \tilde g_*$ exists (under the hypothesis that the flow
does not hit a Landau pole before approaching it). Note that the
dimensionless square mass diverges as $\mu^{2-d}$, but the
dimensionful mass goes to zero for $d<4$. This is not incompatible
with the fact the gluon propagator is finite in the zero momentum
limit, because in the present scheme, the mass is {\it not} defined
through the renormalization condition (\ref{rencond}). It is easy to
show that:
\begin{equation}
\lim_{p\to 0}\Gamma_A^{(2)}(p)=\frac{\tilde g_*^2
  m^4(\mu_0)}{\tilde m_{\text as}^2 g^2(\mu_0)},
\end{equation}
where $\tilde m^2(p)\sim \tilde m_{\text as}^2 p^{2-d}$. The $d=2$
case is different. In that dimension, the coupling constant would
approach a fixed point in an hypothetical IR regime, but the leading
contribution to the flow of $\tilde m^2$ is zero and one must analyze
sub-leading contributions. This important particular case is analyzed
in Sect.~\ref{2dcase}.

\subsection{IR safe scheme in $d=3$}
\label{irsafescheme3d}

Having considered the general IR behavior, we now repeat the calculation of the Sect.~\ref{irsafescheme} in the
$d=3$ case and compare the results with the lattice simulations. The calculation now leads to the following anomalous
dimensions:
\begin{equation}
\begin{split}
\gamma_A=&\frac{g^2 N }{128 \mu  \pi 
   t^2}\Big( \pi t^2 \big(-3 t^2+2 \big)
+4 \sqrt{t} \left(7 t^2-29 t+15\right) \\
&-2 t
   \left(3 t^3-8 t^2+40 t-96\right) \arctan \big(\sqrt{t}/2\big)\\
&+4 \left(3
   t^4-4 t^3+8 t^2-15\right) \arctan\big(\sqrt{t}\big)\Big),
\end{split}
\end{equation}
\begin{equation}
\begin{split}
\gamma_c=&\frac{g^2 N}{32 \mu  \pi  t} \Big(-\pi t^2+2 \sqrt{t} \left(t+3\right)\\
&+2 (t+1)(t-3) \arctan\big(\sqrt{t}\big)\Big),
\end{split}
\end{equation}
where again $t=\mu^2/m^2$. We then deduce the $\beta$ functions for
$g$ and $m^2$ via the non-renormalization theorems
(\ref{eq_def_beta_g}) and (\ref{m2IRscheme}). For some initial
conditions the flow shows a Landau pole, but for the particular values
of couplings that do match with numerical simulations, there is no
Landau pole. The dimensionless running mass $\tilde m(\mu)$ increases when $\mu$ decreases
and we get an IR regime when $\mu \ll m$.  As discussed in the
previous section, this regime is characterized by the $\mu \ll m$ limit
of the function $\beta_{\tilde g}$ that for $d=3$ is
\begin{equation}
\beta_{\tilde g}\sim -\frac{\tilde g}{2}+\frac{N \tilde g^3}{128},
\end{equation}
and the corresponding IR fixed point is
\begin{equation}
 N\,\tilde g_*^2= 64.
\end{equation}
In the Sect.~\ref{IR_suppress} it is discussed why, in fact, these
couplings are not as large as it seems here.  We want here only to
discuss the corresponding consequences for gluon and ghosts
propagators. The RG equations can be integrated numerically and from
them, the propagators can be obtained as discussed in the $d=4$. The
corresponding results are presented in Fig.~\ref{fig3d}. We note that
the inclusion of these RG running considerably improve the results
when compared with those of strict one-loop perturbative results and
now compare very well with lattice simulations. This is not
surprising, because even if the present model does not have IR
divergences it is very usual that RG effects play an important role in
field theories below four dimensions even at moderate values of the
couplings.

\subsection{An estimate of higher loop corrections}
\label{IR_suppress}
The coupling constants that appear along this work are frequently
relatively large at intermediate scales $\mu\sim 1$~GeV. This
situation is much better than having a Landau pole, but it is anyway
problematic for a perturbative analysis. In this section we argue that
although the coupling constant is not small, the perturbation theory
seems to be under control anyway.

\begin{figure}[tbp]
\epsfxsize=7.6cm
\epsfbox{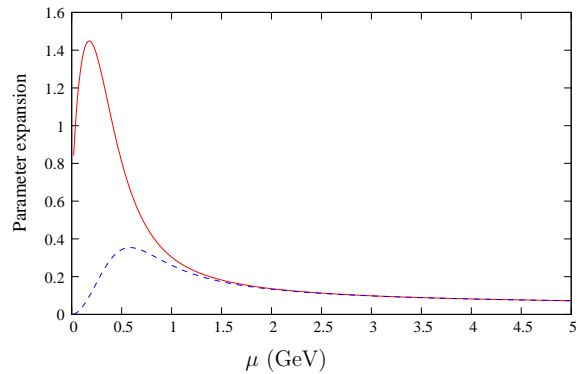}
\caption{\label{4dparam}Four dimensional expansion parameter for
  $SU(3)$ (color online).  Plain curve (red), $u(\mu)$ is the
  dimensionless parameter expansion, valid in the UV. Dashed curve
  (blue), $u(\mu)\mu^2 /(\mu^2+m^2(\mu))$ takes into account the IR
  freeze-out.  The parameters are fixed at 1 GeV as in
  Sect.~\ref{irsafescheme}.}
\end{figure}

First of all, as is well known \cite{weinberg}, the parameter expansion is not
$g^2$ but, because of angular factors and group factors:
\begin{equation}
\label{expansionparam}
u(p)=\frac{g^2(p) Np^{d-4}}{(4\pi)^{d/2}\Gamma(d/2)}.
\end{equation}
where $p$ is the momentum scale used to define the coupling constant.
In $d\neq 4$, $g$ has dimension $p^{(4-d)/2}$ and the power of momenta
in the previous formula is necessary in order to have a dimensionless
expansion parameter.  The angular factors considerably reduces the
size of the expansion parameter. For example, for $d=4$ and $N=3$,
this expansion parameter has a maximum at values of order one as shown
in Fig.~\ref{4dparam}.

There is actually another effect that suppresses the radiative
corrections in the IR as we discuss now in detail.  Note that when
momenta are much smaller than the gluon mass, all diagrams that
include internal gluon lines are suppressed by inverse powers of the gluon
mass. This has two consequences. First, as shown in
Appendix~\ref{appendix_suppress}, the present model does not present
IR divergences for $d>2$ at non-exceptional momenta. This is
surprising at first sight since there are massless modes (ghosts), but
it can be understood because their interactions are mediated by
massive gluons.  A second consequence is that in that regime, the
leading contributions are those with a minimum number of gluon
propagators and we can consider an effective theory where the only
dynamical degrees of freedom are the (massless) ghosts.  In this
effective theory, gluons appear only as external sources coupled to
the ghosts via the bare ghost-gluon vertex while the ghosts interact
via an effective four-point vertex that behaves as $ p_1 p_2 g^2/m^2$,
where $p_1$ and $p_2$ are the anti-ghost momenta
\footnote{We show in the Appendix~\ref{appendix_suppress} that this
  effective field theory is also IR safe for non-exceptional momenta
  for $d>2$.}.  This implies that the effective expansion parameter is
suppressed by $p^2/m^2$ where $p$ is a typical external momentum.  A
naive interpolation between the UV parameter expansion $u$ given in
(\ref{expansionparam}) and the one relevant for the IR would be:
\begin{equation}
 u(p)\frac {p^2}{p^2+m^2}= \frac{g^2(p) N p^{d-2}}{(4\pi)^{d/2}\Gamma(d/2) (m^2+p^2)}.
\end{equation}
As shown in Fig.~\ref{4dparam} this parameter is indeed relatively
small and this makes a perturbative analysis well founded.  Actually,
in a small region around 1~GeV, the expansion parameter is of the
order of 0.4. This means that 2-loop corrections in this region will
give contributions of order $0.4^2$ ($\sim 15\%$), which fits with the
discrepancy between the 1-loop results and the lattice results shown
in Fig.~\ref{fig4drg}. A 2-loop calculation would give a more precise
estimate of higher order corrections.

\begin{figure}[tbp]
\epsfxsize=7.6cm
\epsfbox{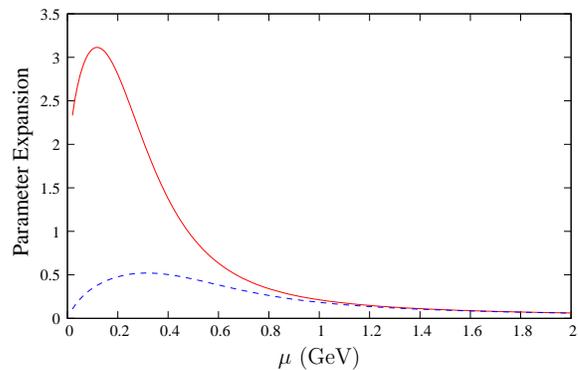}
\caption{\label{3dparam}Three-dimensional expansion parameter for
  $SU(2)$ (color online).  Plain curve (red), $u(\mu)$ is the
  dimensionless parameter expansion, valid in the UV.  Dashed curve
  (blue), $u(\mu)\mu^2 /(\mu^2+m^2(\mu))$ takes into account the IR
  freeze-out.  The parameters are fixed as in
  Sect.~\ref{irsafescheme3d}.}
\end{figure}

Similar results apply for other values of $N$
and $d$. As an example, the same curve is presented for $N=2$ and
$d=3$ in Fig~\ref{3dparam}. It is shown that the actual values in the
IR are slightly larger but remain relatively small.

\section{Discussion of the $d=2$ case}
\label{2dcase}

The explicit expression for propagators can also be obtained in
$d=2$. As before, there are no UV divergences for the
same reasons as for $d=3$. The corresponding result is:
\begin{equation}
\Gamma_{\bar c c}^{(2),\text{1\,loop}}(p)=\frac{g^2 N }{8\pi}(s\log (s) - (s + 1)\log (s + 1)),
\end{equation}
\begin{equation}
\begin{split}
\Gamma_{A}^{(2),\text{1\,loop}}&(p)=\frac{g^2 N}{16\pi s}\Big(- 
          2 (s - 1)^2 (s + 1)\log (s + 1)\\
&-\sqrt{s (s + 4)}\log\left(\frac{\sqrt{s + 4} - \sqrt{s}}{\sqrt{s + 4}+\sqrt{s}}\right) (s - 
                  2)^2 \\&+ s\left(s^2 - 2\right)\log (s) \Big),
\end{split}
\end{equation}
where, again $s=p^2/m^2$.

Within the model, the difference between $d=2$ and $d>2$
which is observed in the lattice (see the introduction) appears natural. 
In $d=2$, we
find that the gluon propagator and ghost dressing function develop logarithmic divergences when $p\to 0$:
\begin{align}
G^{-1}(p)&\sim -\frac{g^2 N}{4\pi}\log \Big(\frac p m \Big), \nonumber\\
F^{-1}(p)&\sim  \frac{g^2 N }{4\pi\, m^2}\log \Big(\frac p m \Big).
\end{align}
Such divergences exclude the possibility of controlling as strict
one-loop calculation as was used above for $d>2$. A proper treatment
of the $d=2$ case requires a RG approach adapted to the IR regime. 

As shown in the appendix \ref{appendix_suppress}, the IR convergence
of the loop integrals in $d=2$ does not improve when increasing the
order of perturbation theory, as it does for $d>2$. This is the
typical situation at the upper critical dimension of a model and it is
well known that a RG treatment is then necessary to study the IR
behavior. We therefore apply the IR safe scheme described in
Sect.~\ref{irsafescheme} in the $d=2$ case.

The anomalous dimensions read:
\begin{align}
\gamma_A&=\frac{g^2 N}{8\pi \mu^2 t}\Big(-t^3 \log t +2 (t^3-4)\log (1+t)\nonumber\\
&+\sqrt{\frac t {4+t}}(t-2)(t^2+4t+12) \log 
\Big(\frac{\sqrt{t+4}-\sqrt{t}}{\sqrt{t+4}+\sqrt{t} }\Big) \Big)
\end{align}
\begin{align}
\gamma_c&=-\frac{N g^2}{4\pi \mu^2 } \log (1+t)
\end{align}
with $t=\mu^2/m^2$. As before, we can extract from them the $\beta$ functions for the
mass and coupling constant. As discussed at the end of
Sect.~\ref{sect_ir}, the leading behavior of the flow of the
dimensionless mass is vanishing in $d=2$. In order to study the IR
behavior, it is convenient to consider the following dimensionless
combination:
\begin{equation}
 \lambda=\frac{g}{m},
\end{equation}
whose $\beta$ function takes the form:
\begin{align*}
 \beta_\lambda&=\frac \lambda 2 \gamma_c\\
&=-\frac{N \lambda^3}{8\pi}\frac{\log(1+t)}t.
\end{align*}
Observe that this $\beta$ function is always negative and therefore
leads to a IR Landau pole. The only way round would be that $\tilde
m^2=1/t$ would run to zero sufficiently fast when $\mu$
decreases. However the study of the flow of $\tilde m^2$ in the IR is
at odds with this hypothesis. Indeed, the $\beta$ function for $\tilde
m^2$ behaves at small $\tilde m^2$ as:
\begin{equation}
\beta_{\tilde m^2}=-\tilde m^2\left(2+\frac{N\lambda^2}\pi\frac{\log (t)}t\right)
\end{equation}
which leads to an increase of $\tilde m^2$ in the IR.

In conclusion, the case $d=2$ is very different from higher
dimensions: There is no fixed point and the flow runs to a Landau
pole. We relate this first property with the observation that, in
$d=2$ and for $\mu\ll\tilde m^2$, the $\beta$ functions for $g$ and
$\tilde m^2$ are proportional (see Eqs.~(\ref{IRbetaanyd}) and
(\ref{eq_m2IR})):
\begin{equation}
2\frac{\beta_{\tilde g}}{\tilde g}=\frac{\beta_{\tilde m^2}}{\tilde m^2}
\end{equation}
A similar situation appears in the random mass Ising model
\cite{Grinstein76} which is characterized by two ``$\phi^4$'' coupling
constants. The associated $\beta$ functions are found to be proportionnal (in the
sense of the previous equation) at leading order in $\epsilon$
expansion. The two-loop calculation \cite{Khmelnitskii76,Jayaprakash77}
lifts this degeneracy and yields a fixed point with coupling constants
$\sim\sqrt\epsilon$. It would be interesting to study if the two-loop
contributions lift in the same manner the degeneracy of the $\beta$ functions for
$\tilde g$ and $\tilde m^2$.

\section{Conclusion}
We have shown in detail that a particular case of the Curci-Ferrari
model, motivated by phenomenological considerations reproduces
quantitatively several non-trivial correlation functions of pure
gluodynamics in all ranges of momenta. The result is obtained from a
very simple one-loop perturbative calculation and the surprising
agreement is justified by estimating the size of higher loop
corrections. Moreover, we have proposed an IR-safe RG scheme that does
not show a Landau pole. Therefore we have at hands a phenomenological
model that correctly describes Yang-Mills correlators. These results
question the common idea that the IR properties of QCD are beyond the
scope of perturbation theory.

At the conceptual level, it would be more satisfying to be able to
calculate the mass parameter from first principles and this is
probably a non-perturbative and very difficult issue (see however
\cite{Verschelde01,Browne03}). However, once the existence of this
mass is accepted, the rest of the analysis seems to be purely
perturbative. The determination of the mass from first principles
would be also interesting since it would fix the results in terms of
the single parameter appearing in pure YM. Another important open
issue is that of the unitarity of the model. A discussed in Sect. II,
usual proofs of unitarity do not work in this model because the
nilponcy of the BRST symmetry is broken. The situation is actually
common to all schemes beyond the FP procedure, such as the GZ model,
but it does not unavoidably mean that the model breaks unitarity. It
could be that a procedure permits to reduce the state space down to a
physical Hilbert space (where unitarity is recovered) that would not
rely on the nilpotency of the BRST operator.  Devising such a
procedure is an open issue.

Beyond these fundamental problems, the evidences that the model
considered here is a good and simple phenomenological one opens the
way to many applications. Let us mention here some of them. 3-point
correlation functions have been measured on the lattice and their
calculation is a direct extension of the work presented here and would
give another test of the ability of the model to reproduce YM
correlation functions. It would also be interesting to compute the
2-loop contributions to the ghost and gluon propagators, that would in
particular give a direct measure of the size of higher order
corrections. The calculation of RG flows and 2-point correlation
functions can be easily extended at finite temperature and many
thermodynamic properties can be extracted from it. Of great physical
interest would be to calculate within this model the quark-antiquark
potential in the quenched approximation, although it is hard to
believe that confinement would appear in such a simple analysis. Let
us mention finally that introducing matter fields is straightforward
and studying the influence of the quarks on the ghost and gluon
correlation functions is a natural extension of this work.

\vspace{.3cm}

{\it Acknowledgments.} We thank A. Cucchieri, O. Oliveira,
M. Mueller-Preussker, A. Sternbeck for useful
correspondence. M.T. thank the IFFI for its
hospitality. N.W. acknowledge the support of the PEDECIBA program.

\vspace{-.3cm}

\appendix

\section{Non-renormalization theorem for the mass}
\label{sect_non-renorm}
We derive in this section the non-renormalization theorem for the
mass. We used this result in Sect.~\ref{zeromomscheme} to simplify the
RG analysis. The main part of the proof is done with bare quantities
but in order to simplify the notations we omit the subscript $B$.

The proof makes use of the Slavnov-Taylor identity and we therefore
introduce sources not only for the primary fields $A_\mu$, $c$,
$\bar c$ and $h$, but also for their BRST variations.
The partition function then reads:
\begin{equation}
e^W=\int\mathcal DA_\mu \mathcal Dc\mathcal D\bar c\mathcal Dh\; e^{-S+S_{\text{sources}}},
\end{equation}
where 
\begin{equation}
\begin{split}
S_{\text{sources}}=\int d^dx (&J_\mu A_\mu+\bar\chi c\\&+\bar c\chi+Rh+\bar
K sA_\mu+\bar L sc),
\end{split}
\end{equation}
and $S$ is the integral of the lagrangian (\ref{eq_lagrang}). The BRST
variations $s$ of the fields are defined as the prefactors of $\eta$
in the right hand sides of (\ref{BRST}). It is not necessary to
introduce sources for the variation of $h$ or for the variations of
variations of the primary fields since these are either vanishing or
linear in the primary fields.  We also introduce the Legendre
transform as:
\begin{equation}
\Gamma+W=\int d^dx\left( J_\mu A_\mu+\bar\chi c+\bar c\chi+Rh\right).
\end{equation}

We first make the change of variables $\bar c(x)\to\bar
c(x)+\bar\eta(x)$. This yelds the equation:
\begin{equation}
\partial_\mu\frac{\delta \Gamma}{\delta\bar K_\mu^a(x)}=\frac{\delta
  \Gamma}{\delta \bar c^a(x)}.
\end{equation}
Deriving this equation with respect to $c^b(y)$ and taking the Fourier
transform (our convention is $\tilde f(p)=\int d^dx\exp(ipx)f(x)$), we
get:
\begin{equation}
\label{eq_appendix_shift}
 \Gamma^{(2)}_{\bar c^ac^b}(p)=-ip_\mu\Gamma^{(2)}_{\bar K_\mu^ac^b}(p).
\end{equation}

We moreover need the Slavnov-Taylor equation, which
reads:
\begin{equation}
\begin{split}
\int d^dx&\Bigg\{\frac{\delta\Gamma}{\delta
  A_\mu^a(x)}\frac{\delta\Gamma}{\delta\bar
  K_\mu^a(x)}+ \frac{\delta\Gamma}{\delta
  c^a(x)}\frac{\delta\Gamma}{\delta\bar L^a(x)}\\&-ih^a(x)\frac{\delta\Gamma}{\delta
  \bar c^a(x)}+im^2\frac{\delta\Gamma}{\delta
  h^a(x)}c^a(x)\Bigg\}=0.
\end{split}
\end{equation}
We derive this expression with respect to $A_\nu^b(y)$ and $c^c(z)$
and take this expression at vanishing sources. Using the fact that the
vacuum does not break the ghost number conservation, only two terms
contribute and making the Fourier transform we get:
\begin{equation}
\begin{split}
\label{eq_appendix_gam2}
\Gamma^{(2)}_{A^b_\nu A^a_\mu}(p)\Gamma^{(2)}_{\bar K^a_\mu c^c}(p)&=im^2\Gamma^{(2)}_{A^b_\nu h^c}(p)\\
&=-im^2p_\nu\delta^{bc},
\end{split}
\end{equation}
where we used the fact that the $h$ sector is not renormalized.

Now, using the fact that $\Gamma^{(2)}_{A_\mu^aA_\nu^b}(p)$ is analytic at low
momentum and behaves as $\delta^{ab}\delta_{\mu\nu}G^{-1}(0)$, and
contracting Eq.~(\ref{eq_appendix_gam2}) with $p_\nu$, we obtain at
low energy:
\begin{equation}
G_B^{-1}(0)F_B^{-1}(0)=m_B^2,
\end{equation}
where we have introduced the subscript $B$ to recall that all this
calculation was done (as emphasized at the beginning of the appendix)
with bare quantities.

\section{Infrared behavior of the Feynman diagrams}
\label{appendix_suppress}
In this appendix we study the IR corrections coming from higher order
diagrams.  Our aim is twofold. First we check that the
regular IR behavior found at 1 loop is not modified by higher order
corrections. Second we study the IR behavior of the low energy
effective theory for the ghosts obtained by considering only a minimal
number of gluon propagators (see Sect.~\ref{IR_suppress}).

Consider first a generic diagram of the model (\ref{eq_lagrang}). It
has $N_A$ external gluon legs, $N_c$ external ghost/antighost legs,
$v_3$ three-gluon vertices, $v_4$ four-gluon vertices and $v_c$
ghost-antighost-gluon vertices. Looking at the bare vertex, we see
that each external anti-ghost leg comes with an external momentum.  In
the following when we speak of diagrams, it is understood that these
trivial external momenta dependence have been factorized.

Let us calculate the superficial degree of IR divergence $\omega$ of
this diagram for vanishing external momenta, which characterize the IR
behavior of the associated integrals. A negative superficial degree
of IR divergence indicates that the corresponding diagram is IR
divergent (at zero momenta). Each ghost propagator contributes -2 but
the gluon propagators being massive do not contribute. The 3-point
interactions being proportional to a momentum contribute +1 and each
loop contributes $d$.  Using textbook techniques \cite{itzykson}, it is
easy to prove that: 
\begin{equation}
 \omega\geq d-\frac d 2 N_A-\frac{d-1}{2} N_c+ \frac{d}{2} v_3+ d \,v_4+
 \frac{d-2}{2} v_c.
\end{equation}
Observing that the pre-factors of $v_3$, $v_4$ and $v_c$ are positive
in $d>2$, we conclude that at fixed number of external legs,
increasing the number of vertices suppress the contributions in the
IR.  Moreover we check that the two-point vertex functions at zero
momenta are finite at all loops. The case $d\leq 2$ is different
because increasing the number of ghost-gluon vertices does not improve
(or even worsen for $d<2$) the IR behavior. In this sense, $d=2$ plays
the role of an upper critical dimension.

The situation is, as expected, more favorable if we consider
non-exceptional momentum configurations.  Using the method described
in \cite{itzykson}, we find that the degree of IR divergence is:
\begin{equation}
 \omega \geq \frac d 2 N_A+\frac{d-1}{2} N_c+ \frac{d+2}{2} v_3+ d \,v_4+
 \frac{d-2}{2} v_c,
\end{equation}
where now $N_A$ and $N_c$ are the external soft legs of the diagram
and $v_3$, $v_4$ and $v_c$ are the number of vertices attached only to
soft momenta. This expression is always positive for $d>2$ but again
$d=2$ plays the role of an upper critical dimension.

We now come to the discussion of the IR behavior of the diagrams in
the effective theory. We consider first diagrams with vanishing
external momenta (as before, we first factorize the external momenta
of the anti-ghosts).  As discussed in Sect.~\ref{IR_suppress}, the
ghosts interact via a 4-point vertex which behaves as the product of
momenta of the anti-ghosts. Denoting $v_{4c}$ the number of such
vertices in a diagram, we find the IR degree of divergence:
\begin{equation}
 \omega=d- N_A-\frac{d-1}{2} N_c+(d-2) v_{4c}.
\end{equation}
Again, we see that, at fixed number of external legs, increasing the
number of vertices suppress the contributions in the IR and check that
the two-point vertex functions at zero momenta remain finite at higher
loops.  Using non-exceptional momentum configurations, we get:
\begin{equation}
 \omega=\frac{d-1}{2} N_c+(d-2) v_{4c},
\end{equation}
which ensures that the diagrams are IR finite for $d\geq 2$.

\end{document}